\documentclass[12pt]{iopart}
\pdfoutput=1
\usepackage[utf8, latin1]{inputenc}

\eqnobysec

\usepackage{iopams}
\usepackage[varg]{txfonts}

\usepackage[plainpages=false]{hyperref}
\usepackage{xparse}

\newcommand{\ie}{i.\,e.~}

\newcommand{\eg}{e.\,g.~}

\newcommand{\wrt}{w.\,r.\,t.~}

\newcommand{\ii}{\mathrm{i}}
\newcommand{\ff}{\mathrm{f}}
\newcommand{\qq}{\mathrm{q}}
\newcommand{\pp}{\mathrm{p}}
\newcommand{\dd}{\mathrm{d}}

\newcommand{\II}{\mathrm{I}}
\newcommand{\CC}{\mathrm{C}}
\newcommand{\GC}{\mathrm{GC}}
\newcommand{\g}{\mathrm{g}}

\newcommand{\gqq}{\g_{\mathrm{qq}}}
\newcommand{\gqp}{\g_{\mathrm{qp}}}
\newcommand{\gpq}{\g_{\mathrm{pq}}}
\newcommand{\gpp}{\g_{\mathrm{pp}}}

\newcommand{\ve}{\vec{e}}

\newcommand{\xin}{x^{(\mathrm{i})}}
\newcommand{\vx}{\vec{x}}
\newcommand{\vxi}{\vx^{\,(\mathrm{i})}}
\newcommand{\tx}{\boldsymbol{x}}
\newcommand{\txi}{\tx^{(\mathrm{i})}}
\newcommand{\btx}{\boldsymbol{\bar{x}}}

\newcommand{\qin}{q^{(\mathrm{i})}}
\newcommand{\vq}{\vec{q}}
\newcommand{\vqi}{\vq^{\,(\mathrm{i})}}
\newcommand{\tq}{\boldsymbol{q}}
\newcommand{\tqi}{\tq^{(\mathrm{i})}}

\newcommand{\pin}{p^{(\mathrm{i})}}
\newcommand{\vp}{\vec{p}}
\newcommand{\vpi}{\vp^{\,(\mathrm{i})}}
\newcommand{\tp}{\boldsymbol{p}}
\newcommand{\tpi}{\tp^{(\mathrm{i})}}

\newcommand{\vs}{\vec{s}}

\newcommand{\kin}{k^{(\mathrm{i})}}
\newcommand{\vk}{\vec{k}}
\newcommand{\vki}{\vk^{\,(\mathrm{i})}}

\newcommand{\vl}{\vec{l}}

\newcommand{\vchi}{\vec{\chi}}
\newcommand{\vochi}{\hat{\vchi}}
\newcommand{\tchi}{\boldsymbol{\chi}}

\newcommand{\T}{\top}						
\newcommand{\mpd}{\bar{\rho}} 						
\newcommand{\dirac}{\delta_{\textsc{d}}}				
\newcommand{\heavi}{\Theta}						

\newcommand{\Diag}[1]{\Sigma_\mathrm{Diag}^{(#1)}}				
\newcommand{\CDiag}[1]{\Sigma_\mathrm{CDiag}^{(#1)}}				
\newcommand{\sfcd}[1]{\tilde{\Sigma}_{\mathrm{CDiag}}^{(#1)}}		

\newcommand{\CorPol}{\hat{\mathcal{C}}}					
\newcommand{\CorOp}{\hat{\mathcal{C}}_{\mathrm{tot}}}			

\newcommand{\cgf}[1]{W_0^{(#1)}}					
\newcommand{\unavZ}[1]{\tilde{Z}_0^{(#1)}}

\newcommand{\tJ}{\boldsymbol{J}}					
\newcommand{\tK}{\boldsymbol{K}}					

\newcommand{\PhiOp}[1]{\hat{\Phi}^{(#1)}}				

\NewDocumentCommand \Gf{o}{						
  \IfNoValueTF{#1}
    { G^{(0)} }   
  { G^{(0,#1)} }
  }

\newcommand{\tL}{\boldsymbol{L}}					
\newcommand{\Lq}[1]{\vec{L}_{\mathrm{q},\mathrm{I}_{#1}}}		
\newcommand{\Lp}[1]{\vec{L}_{\mathrm{p},\mathrm{I}_{#1}}}		

\newcommand{\Ki}{\vec{\mathcal{K}}^{\,(\mathrm{i})}}			

\newcommand{\upi}[1]{\int \mathcal{D} #1 \,}				

\newcommand{\bpint}[3]{\int\limits_{#2}^{#3} \mathcal{D} #1 \,} 	

\newcommand{\usi}[1]{\int \mathrm{d} #1 \,}				

\newcommand{\bsi}[3]{\int\limits_{#2}^{#3} \mathrm{d} #1 \,}		

\newcommand{\umi}[2]{\int \mathrm{d}^{#1} #2 \,}			

\newcommand{\fmi}[2]{\int \frac{\mathrm{d}^{#1} #2}{(2\pi)^{#1}} \,}    

\newcommand{\IC}{\int \mathrm{d} \Gamma_\ii \,}				

\newcommand{\fd}[2]{\frac{\delta #1}{\ii \delta #2}}			

\newcommand{\ave}[1]{\left\langle #1 \right\rangle}			

\newcommand{\nexp}[1]{\mathrm{exp}\left\{ #1 \right\}}			

\newcommand{\tens}[1]{\boldsymbol{#1}} 					

\newcommand{\cvector}[1]{\left(\begin{array}{c}#1\end{array}\right)}	

\renewcommand{\matrix}[2]{\left(\begin{array}{#1} #2\end{array}\right)}  

\usepackage{tikz}

\usetikzlibrary{
positioning,
decorations.pathmorphing,
decorations.pathreplacing,
decorations.markings,
shapes.geometric,
calc
}

\newlength{\gs}
\newlength{\pdotdiam}
\NewDocumentCommand{\mytikz}{O{-0.6ex}}{\tikz[baseline=#1,radius=\gs]}

\def\loop{ \draw[fill=white] (0,0) circle [radius=3pt] ; }

\tikzset{
pdot/.style={draw, black, circle, fill, minimum size=\pdotdiam, inner sep=0},
dline/.style={draw, black, line cap=round, thick},
pline/.style={draw, black, line cap=round, thick, dash pattern=on 2pt off 2pt},
gline/.style={draw, black, line cap=round, thick, dotted},
lline/.style={draw, black, line cap=round, thick, postaction={ decorate, decoration={markings, mark=at position 0.5 with {\loop}} } }
}

\NewDocumentCommand \pdot{O{}}{
\coordinate[pdot] (p1) at (0,0);
\node[below] at (p1) {$#1$};
}

\NewDocumentCommand \hortwopattern{O{}O{}}{
\coordinate[pdot] (p1) at (0,0);
\node [below] at (p1) {$#1$};
\coordinate[pdot] (p2) at (\gs,0);
\node [below] at (p2) {$#2$};
}

\NewDocumentCommand \vertwopattern{O{}O{}}{
\coordinate[pdot] (p1) at (0,0);
\node [below] at (p1) {$#1$};
\coordinate[pdot] (p2) at (0,\gs);
\node [above] at (p2) {$#2$};
}

\NewDocumentCommand \threepattern{O{}O{}O{}}{
\coordinate[pdot] (p1) at (0,0);
\node [below] at (p1) {$#1$};
\coordinate[pdot] (p2) at (\gs,0);
\node [below] at (p2) {$#2$};
\coordinate[pdot] (p3) at (0.5\gs,\gs);
\node [above] at (p3) {$#3$};
}

\NewDocumentCommand \fourpattern{O{}O{}O{}O{}}{
\coordinate[pdot] (p1) at (0,0);
\node [below] at (p1) {$#1$};
\coordinate[pdot] (p2) at (\gs,0);
\node [below] at (p2) {$#2$};
\coordinate[pdot] (p3) at (\gs,\gs);
\node [above] at (p3) {$#3$};
\coordinate[pdot] (p4) at (0,\gs);
\node [above] at (p4) {$#4$};
}

\newlength{\edgelength}
\NewDocumentCommand \largefourpattern{O{}O{}O{}O{}}{
\coordinate[pdot] (p1) at (0,0);
\node [below left] at (p1) {$#1$};
\coordinate[pdot] (p2) at (\edgelength,0);
\node [below right] at (p2) {$#2$};
\coordinate[pdot] (p3) at (\edgelength,\edgelength);
\node [above right] at (p3) {$#3$};
\coordinate[pdot] (p4) at (0,\edgelength);
\node [above left] at (p4) {$#4$};
}

\NewDocumentCommand \ddline{mmo}{
  \IfNoValueTF{#3}
  {\draw[dline] (#1) -- (#2);}
  {\draw[dline] (#1) .. controls ($($(#1)!0.25!(#2)$)!#3!90:(#2)$) and ($($(#1)!0.75!(#2)$)!#3!90:(#2)$) .. (#2);}
}

\NewDocumentCommand \ppline{mmo}{
  \IfNoValueTF{#3}
  {\draw[pline] (#1) -- (#2) ; }
  {\draw[pline] (#1) .. controls ($($(#1)!0.25!(#2)$)!#3!90:(#2)$) and ($($(#1)!0.75!(#2)$)!#3!90:(#2)$) .. (#2);}
}

\NewDocumentCommand \ggline{mmo}{
  \IfNoValueTF{#3}
  {\draw[gline] (#1) -- (#2) ; }
  {\draw[gline] (#1) .. controls ($($(#1)!0.25!(#2)$)!#3!90:(#2)$) and ($($(#1)!0.75!(#2)$)!#3!90:(#2)$) .. (#2);}
}

\NewDocumentCommand \dpline{mmo}{
  \IfNoValueTF{#3}
  { \draw[dline] (#1) -- ($(#1)!0.45!(#2)$) ;
    \draw[pline] ($(#1)!0.45!(#2)$) -- (#2) ;
  }
  {
    \begin{scope}
     \clip (#1) rectangle ($($(#1)!0.5!(#2)$)!#3!90:(#2)$);    
     \draw[dline] (#1) .. controls ($($(#1)!0.25!(#2)$)!#3!90:(#2)$) and ($($(#1)!0.75!(#2)$)!#3!90:(#2)$) .. (#2);
    \end{scope}
    \begin{scope}
     \clip (#2) rectangle ($($(#1)!0.5!(#2)$)!#3!90:(#2)$);
     \draw[pline] (#1) .. controls ($($(#1)!0.25!(#2)$)!#3!90:(#2)$) and ($($(#1)!0.75!(#2)$)!#3!90:(#2)$) .. (#2);
    \end{scope}
  }
}

\NewDocumentCommand \ddlline{mmo}{
  \IfNoValueTF{#3}
  {\draw[lline] (#1) -- (#2);}
  {\draw[lline] (#1) .. controls ($($(#1)!0.25!(#2)$)!#3!90:(#2)$) and ($($(#1)!0.75!(#2)$)!#3!90:(#2)$) .. (#2);}
}

\NewDocumentCommand \arrline{mm}{
 \draw[dline,->] (#1) -- ($(#1)!0.55!(#2)$);
 \draw[dline] ($(#1)!0.55!(#2)$) --(#2);
}

\begin{document}

\title{Kinetic Field Theory: Exact free evolution of Gaussian phase-space correlations}
\author{Felix Fabis, Elena Kozlikin, Robert Lilow, Matthias Bartelmann}
\address{Heidelberg University, Zentrum f\"ur Astronomie, Institut f\"ur Theoretische Astrophysik, Philosophenweg 12, 69120 Heidelberg, Germany}

\begin{abstract}
In recent work we developed a description of cosmic large scale structure formation in terms of non-equilibrium ensembles of classical particles, with
time evolution obtained in the framework of a statistical field theory. In these works, the initial Gaussian correlations between particles have so far been treated perturbatively or restricted
to pure momentum correlations. Here we treat the correlations between all phase-space coordinates exactly by adopting a diagrammatic language for the different forms of correlations,
directly inspired by the Mayer cluster expansion. We will demonstrate that explicit expressions for phase-space density cumulants of arbitrary $n$-point order, which fully capture the non-linear
coupling of free streaming kinematics due to initial correlations, can be obtained from a simple set of Feynman rules.
These cumulants will be the foundation for further investigations of interacting perturbation theory.
\end{abstract}


\maketitle

\section{Introduction} \label{sec:introduction}

Based on the works of Das \& Mazenko \cite{Mazenko2010,Mazenko2011,Das2012,Das2013} we have developed a variant of kinetic theory \cite{Bartelmann2016,Bartelmann2017,Viermann2015}
we have dubbed `Kinetic Field Theory' (KFT). Our primary goal is to find a new approach to cosmic structure formation which is free from the limitations of established methods
like Eulerian standard perturbation theory (SPT). For a discussion of the advantages we refer to our previous work and announce an upcoming paper comparing KFT and SPT directly in more detail.

The central idea of KFT is to use the path integral approach to classical mechanics (cf.~\cite{Martin1973,Gozzi1989,Penco2006}) in order to directly encode the microscopic phase-space dynamics of
individual particles in a generating functional. Collective macroscopic fields are then constructed from the microscopic information at the time of interest.
We adopted this idea in order to describe the time evolution of the non-equilibrium statistics of a system of gravitationally interacting particles
whose initial positions and momenta are correlated in such a way that they sample Gaussian density and velocity fields.

In \cite{Bartelmann2016} we employed perturbation theory in the interaction potential relative to a modified version of Zel'dovich trajectories \cite{Bartelmann2015a}.
This reproduced the non-linear growth of the CDM density fluctuation power spectrum known from $N$-body simulations over a wide range of scales remarkably well considering we included perturbative
corrections only to first order. In this first work the initial correlations were also perturbatively expanded into orders of the initial density fluctuation power spectrum.
This allowed straightforward numerical evaluation but restricted the validity of even the free theory to certain time and length scales and made the
formalism appear less elegant than it actually is.

Consequently, we layed out a more general approach in \cite{Bartelmann2017} focusing on an ensemble of particles correlated in their momenta only. By taking these correlations into account exactly
and artificially reducing damping effects (see \cite{Dombrowski2017} for a motivation of this workaround) we could show that a substantial part of the late-time non-linear growth
behaviour of the density fluctuation power spectrum is actually already encoded in the initial correlations. This further corroborates the hope that low-order perturbation theory
relative to Zel'dovich-type trajectories will yield accurate results for the evolution of density statistics.

Restricting initial correlations to the momenta only is an accurate description of cosmic structure formation at late times when using a particle propagator which is not bound
from above like in the case of Zel'dovich-type trajectories, because momentum correlations scale with higher polynomial orders of the propagator compared to density correlations.
However, when using a strict separation between non-interacting kinematic motion and interactions the particle propagator will be bound from above \cite{Bartelmann2015a} due to cosmic expansion.
Considering this case will be necessary for a direct comparison of the non-interacting evolution of density statistics of both KFT and SPT.

More importantly, we will need the cumulants including the exact initial correlations between all phase-space coordinates as the building blocks for an alternative formulation of perturbation
theory built around macroscopic fields while still retaining the underlying particle dynamics. This alternative approach will also be developed in a separate paper.
It will naturally lead to a partial resummation of the canonical perturbation theory in \cite{Bartelmann2016}. In order for this resummed theory to reproduce the familiar linear growth
behaviour of the matter power spectrum known from SPT, one needs to include density and momentum correlations on equal footing.

In the main body of this work we concentrate on the conceptual steps leading to our central result, \ie the general form of non-interacting collective field cumulants
in terms of diagrams evaluated according to Feynman rules. Detailed calculations and derivations can be found in the Appendices.
Since our results only depend on very general properties of the force acting between particles, our results in the present work should in principle
have a much broader range of applicability than just cosmic structure formation. We will thus endeavour to keep the mathematical formulation of our results as general as possible, even
though we will most often try to make connections to cosmology when commenting on them.

In Section \ref{sec:gen_func} we give a short recapitulation of the KFT approach by setting up the generating functional for a particle ensemble described initially
by a Gaussian random field of density and momentum.
The focus of Section \ref{sec:cluster_expansion} lies on re-ordering the free generating functional in a way that reflects the particle-based nature of the theory more clearly.
The main result will be a grand canonical version of the generating functional in the form of a sum over connected diagrams representing initial particle correlations
ordered by the number of correlated particles.
In Sections \ref{sec:density_cumulants} and \ref{sec:mixed_cumulants} we explain the calculation of exact and explicit expressions for non-interacting $n$-point cumulants of the collective fields.
We extend the results of \cite{Bartelmann2016} from the spatial density $\rho$ to the full phase-space density $f$.
Their calculation essentially reduces to simple Feynman rules for the initial correlation diagrams.
Finally, Section \ref{sec:conclusions} will contain a summary of the paper and present our conclusions.

\subsection{Notation} \label{sec:notation}

We will follow the notation of \cite{Bartelmann2016} for the microscopic phase-space coordinates
of a large set of $N$ particles confined to a volume $V$. Individual particles are enumerated with \textit{particle labels} $j=1,\ldots,N$ and have positions and momenta denoted as
$d$-dimensional vectors $\vq_j$ and $\vp_j$. We combine those into a $2d$-dimensional phase-space coordinate vector as
\begin{equation}
 \vx_j = \cvector{\vec{q}_j \\ \vp_j} \quad \forall j = 1,\dots,N \;.
\label{eq:phase_space_coordinates}
\end{equation}
For all $N$ particles we bundle these vectors with the help of the tensor product
\begin{equation}
 \tens{q} = \vq_j \otimes \ve_j \;, \qquad \tp = \vp_j \otimes \ve_j \;, \qquad \tx = \vx_j \otimes \ve_j \;,
\label{eq:phase_space_tensor}
\end{equation}
where $\ve_j$ is the Cartesian base vector in $N$ dimensions with entries $(\ve_j)_i = \delta_{ij}$. The Einstein summation convention is implied unless explicitly stated otherwise or obvious from context.
The bold vectors from \eref{eq:phase_space_tensor} follow the rules of matrix multiplication inducing a scalar product 
\begin{eqnarray}
 \tens{a} \cdot \tens{b} 
 &=\tens{a}^{\T} \, \tens{b} 
 = \left(\vec{a}^{\,\T}_j \otimes \vec{e}^{\,\T}_j\right) \left(\vec{b}_k \otimes \vec{e}_k\right) 
 = \vec{a}^{\,\T}_j \, \vec{b}_k \, \delta_{jk}
 = \vec{a}_j \cdot \vec{b}_j \;.
\label{eq:tensor_dot_product}
\end{eqnarray}
For functions of time we extend the dot product to include an implied time integration as
\begin{equation}
 \tens{a} \cdot \tens{b} = \bsi{t}{t_\ii}{t_\ff}\tens{a}(t) \cdot \tens{b}(t) \;.
\label{eq:tensor_dot_product_time_integration}
\end{equation}
We will also encounter tuples of fields defined on the domains of time and space. For two tuples $A,B$, the dot product additionally includes an implied integration over Fourier space as
\begin{equation}
 A \cdot B = \usi{r} A^{\T}(-r) \, B(r) = \fmi{2d}{s_r} \bsi{t_r}{t_\ii}{t_\ff} A^{\T}(-r) \, B(r) \;,
\label{eq:fields_dot_product}
\end{equation}
which we extend in an analogous fashion to matrix products. We have abbreviated field arguments as \textit{field labels} $(\pm r) \coloneqq (\pm \vec{s}_r, t_r) = (\pm \vec{k}_r, \pm \vec{l}_r, t_r)$,
where $\vs$ is Fourier-conjugate to $\vx$, $\vk$ to $\vq$ and $\vl$ to $\vp$. 
We also denote integration over phase-space and its conjugate Fourier space by
\begin{eqnarray}
 \int_x \coloneqq \umi{2d}{x} = \umi{d}{q} \umi{d}{p} = \int_q \int_p \;, \nonumber \\
 \int_s \coloneqq \fmi{2d}{s} = \fmi{d}{k} \fmi{d}{l} = \int_k \int_l \;.
\label{eq:integrals_definition}
\end{eqnarray}

\section{The generating functional} \label{sec:gen_func}

\subsection{Kinetic field theory} \label{sec:kft}

We will only briefly go through the main aspects of KFT, for a more detailed introduction see \cite{Bartelmann2016} or alternatively the original works by Das \& Mazenko in
\cite{Mazenko2010,Mazenko2011,Das2012,Das2013}. 
As already mentioned in the introduction, the central idea of the KFT approach is to encapsulate both the dynamics and the initial statistics of
a non-equilibrium $N$-particle system in a generating functional. Time evolution is accounted for by functionally integrating over all possible trajectories 
but forcing them to the classical solution of the equations of motion by means of a functional Dirac delta distribution. Stochasticity is introduced into the systems
by averaging over the initial conditions with some initial phase-space probability distribution $\mathcal{P}(\tens{x}^{(\ii)})$. The canonical generating functional thus reads
\begin{equation}
 Z_{\CC} = \usi{\txi} \mathcal{P}(\txi) \bpint{\tx(t)}{\txi}{} \upi{\tchi(t)} \, \e^{ \ii \, \tchi \cdot \tens{E}[\tens{x}] } \;,
\label{eq:gen_func_definition}
\end{equation}
where $\tens{E}[\tx(t)] = 0$ is the equation of motion for the $N$-particle system. We have rewritten the functional Dirac delta distribution as a Fourier transform by introducing
new auxiliary fields $\tchi(t)$. They play the role of the functional Fourier conjugates to the phase-space coordinates $\tx(t)$.
We will deal with systems with Hamiltonian equations of motion for which we assume the form
\begin{equation}
 \tens{E}[\tx] = \left(\partial_t + \tens{F}\right) \tx + \tens{\nabla}_q \, \mathcal{V} = 0 \;, \quad \tens{F} = F_j \otimes \ve_j \;, \quad F_j \vx_j = \cvector{ -\nabla_{p_j} \\ \nabla_{q_j} } \mathcal{H}_0 \;,
\label{eq:equation_of_motion}
\end{equation}
\ie the matrices $F_j$ encapsulate the free part of the equation of motion for individual particles and $\mathcal{V}(\vec{q},t)$ is the interaction potential.
In order to separate free motion from interactions we first introduce generator fields $\vec{J}_j, \vec{K}_j$ paired with $\vx_j, \vchi_j$ respectively.
Under the path integral, we can then replace
\begin{equation}
 \vx_j(t) \rightarrow \hat{\vx_j}(t) \coloneqq \fd{}{\vec{J}_j(t)} \quad \textrm{and} \quad \vchi_j(t) \rightarrow \vochi_j(t) \coloneqq \fd{}{\vec{K}_j(t)} \;.
\label{eq:single_particle_generator_operators}
\end{equation}
Any function of these operators may now be pulled in front of the integrals.

We restrict ourselves to systems of $N$ identical particles in the absence of external forces and assume
that the interaction potential between particles will only depend on the distance between their respective positions at the same point in time, implying that $\mathcal{V}$
may be written as a superposition of $N$ single-particle potentials $v$. This naturally leads to the introduction \cite{Bartelmann2016,Viermann2015} of two
fundamental collective fields $\Phi(r) = (\Phi_f(r), \Phi_B(r))^{\T}$. The Klimontovich phase-space density is then defined as
\begin{equation}
  \Phi_f(r) = \int_{x_r} e^{-\ii\, \vs_r \cdot \vx_r} \sum_{j=1}^N \dirac\left(\vx_r - \vx_j(t_r)\right) = \sum_{j=1}^N e^{-\ii \, \vs_r \cdot \vx_j(t_r)} \;.
\label{eq:phase_space_density}
\end{equation}
It encodes the instantaneous occupation of a phase-space state $\vx$ by the particles of the ensemble, which in the conjugate Fourier space turns into a superposition of phase-factors.
According to \eref{eq:single_particle_generator_operators}, its operator version is
\begin{equation}
 \hat{\Phi}_f(r) = \sum_{j=1}^N \nexp{-\ii \, \vs_r \cdot \fd{}{\vec{J}_j(t_r)}} \eqqcolon \sum_{j=1}^N \hat{\Phi}^{(1)}_{f_j}(r) \;.
\label{eq:phase_space_density_operator}
\end{equation}
The deviation of all individual particles from their inertial trajectories due to the interaction with all other particles is encoded in the response field
\begin{equation}
  \Phi_B(r) = \int_{x_r} e^{-\ii\, \vs_r \cdot \vx_r} \sum_{j=1}^N \vchi_{p_j}(t) \cdot \nabla_q \dirac\left(\vx_r - \vx_j(t_r)\right) = \sum_{j=1}^N \ii \, \vchi_{p_j} \cdot \vk_r \, e^{-\ii \, \vs_r \cdot \vx_j(t_r)} \;.
\label{eq:response_field}
\end{equation}
Again using \eref{eq:single_particle_generator_operators} we obtain its operator version
\begin{equation}
  \hat{\Phi}_B(r) = \sum_{j=1}^N \left( \ii \, \vk_r \cdot \fd{}{\vec{K}_{p_j}(t_r)} \right) \, \hat{\Phi}^{(1)}_f(r) \eqqcolon \sum_{j=1}^N \hat{b}_{j}(r) \, \hat{\Phi}^{(1)}_{f_j}(r) = \sum_{j=1}^N \hat{\Phi}^{(1)}_{B_j}(r) \;.
\label{eq:response_field_operator}
\end{equation}
The interaction potential is encapsulated in the matrix
\begin{equation}
 \sigma(r,r') = - \frac{1}{2} \, \dirac(t_r - t_{r'}) (2\pi)^{3d} \, \dirac\left(\vl_r\right) \dirac\left(\vl_{r'}\right) \dirac\left(\vk_r + \vk_{r'}\right) v(k_r) \matrix{cc}{ 0 & 1 \\ 1 & 0} \;.
\label{eq:interaction_matrix}
\end{equation}
It will be advantageous to introduce source fields $H_f, H_B$ paired with the collective fields $\Phi_f, \Phi_B$. The appropriate source term in Fourier space is given by
\begin{eqnarray}
 \e^{\ii H \cdot \Phi} = \nexp{ \ii \usi{\bar{r}} H(-\bar{r}) \cdot \Phi(\bar{r}) } \; \rightarrow \; \hat{H}_{\alpha(r)} = (2\pi)^{2d} \fd{}{H_\alpha(-r)} \;,
\label{eq:collective_source_term}
\end{eqnarray}
where $\alpha = f,B$ and the hatted $\hat{H}$ operators allow to generate arbitrary factors of $\Phi$ by repeated application.
Following \cite{Bartelmann2016,Viermann2015} the generating functional may then be written as
\begin{eqnarray}
  Z_{\CC}[H,\tens{J},\tens{K}] &= \e^{\ii \, \hat{H} \cdot \sigma \cdot \hat{H} } \e^{\ii \, H \cdot \hat{\Phi} } \IC \bpint{\tx}{\txi}{} \upi{\tchi} \e^{ \ii \left( \tchi \cdot \left(\partial_t + \tens{F}\right) \tx + \tJ \cdot \tx + \tK \cdot \tchi \right) } \nonumber \\
  &= \e^{\ii \, \hat{H} \cdot \sigma \cdot \hat{H} } \, \e^{\ii \, H \cdot \hat{\Phi} } \, Z_{\CC,0}[\tJ,\tK] \;,
\label{eq:gen_func_definition_explicit}
\end{eqnarray}
where we abbreviated $\dd \Gamma_\ii = \usi{\txi} \mathcal{P}(\txi)$. With the generating functional in this form, cumulants of the collective fields are generated
by taking appropriate functional derivatives and then setting all sources to zero,
\begin{equation}
 G_{\alpha_1(1) \dots \alpha_n(n)} \coloneqq \ave{\Phi_{\alpha_1(1)} \dots \Phi_{\alpha_n(n)}}_{\mathrm{c}} = \hat{H}_{\alpha_1(1)} \dots \hat{H}_{\alpha_n(n)} \, \ln Z_{\CC}[H] \, \Big|_{0} \;.
\label{eq:cumulants_definition}
\end{equation}

While the particle interactions will not be important for the remainder of this work, we include them in the appropriate places in order to show that they do not interfere with any of our calculations.
In the case of \eref{eq:gen_func_definition_explicit} it is important to note that we were able to pull the interaction operator in front of the integration over the initial conditions which means the
latter can be performed before any interactions need to be considered. 

We will mostly concern ourselves with the free generating functional $Z_{\CC,0}[\tJ,\tK]$. As shown in \cite{Bartelmann2016}, the path integrals defining it can be executed once
the Green's function $\mathcal{G}$ of the free equation of motion of a single particle is known. One finds
\begin{equation}
 Z_{\CC,0}[\tJ,\tK] = \usi{\txi} \mathcal{P}(\txi) \, \e^{ \ii \, \tJ \cdot \btx }
\label{eq:free_gen_func}
\end{equation}
where the solution $\btx(t)$ to the free and linear equations of motion is given by
\begin{equation}
 \btx(t) = \tens{\mathcal{G}}(t,t_\ii) \, \txi - \bsi{t'}{t_\ii}{t_\ff} \tens{\mathcal{G}}(t,t') \tK(t') \;,
\label{eq:trajectories_free_solution}
\end{equation}
and the non-interacting particle propagator has the general form
\begin{equation}
 \tens{\mathcal{G}}(t,t') = \mathcal{G}(t,t') \otimes \mathcal{I}_N \quad \textrm{with} \quad \mathcal{G}(t,t') = \matrix{cc}{ \gqq(t,t') \mathcal{I}_d & \gqp(t,t') \mathcal{I}_d \\ \gpq(t,t') \mathcal{I}_d & \gpp(t,t') \mathcal{I}_d } \;.
\label{eq:propagator_definition}
\end{equation}
The appearance of the additional source term $\tK$ in the free solution allows for the generation of deviations from the inertial motion of particles. Since the auxiliary fields 
$\vchi_j$ appear in the response field \eref{eq:response_field} and are replaced according to \eref{eq:response_field_operator}, we now explicitly see how deviations from trajectories
couple to interactions with the potential and thus form the \textit{actual} perturbed physical quantity in KFT perturbation theory.

\subsection{Initial phase-space probability distribution} \label{sec:initial_distribution}

Both the theory of inflation and observations of the cosmic microwave background imply that the initial cosmic density $\rho^{(\ii)}(\vec{q})$ 
and momentum $\vec{P}^{\,(\ii)}(\vec{q})$ fields at the beginning of the matter dominated epoch should together constitute a Gaussian random field.
Apart from its importance in cosmology, this initial condition is also an instructive example for the simplest possibility to include correlations,
since any Gaussian field is completely defined by its mean and its covariance matrix. Due to the cosmological principle it must be statistically homogeneous and isotropic, implying
\begin{equation}
  \ave{\cvector{\rho^{(\ii)}(\vec{q}) \\ \vec{P}^{\,(\ii)}(\vec{q})}} = \cvector{\mpd^{(\ii)} \\ \vec{0}} \;, \quad \rho^{(\ii)}(\vec{q}) \eqqcolon \mpd^{(\ii)} (1 + \delta^{(\ii)}(\vec{q})) \;,
\label{eq:density_contrast_definition}
\end{equation}
where the average is taken over the Gaussian distribution of the field values. The homogeneous mean particle density is given by $\mpd^{(\ii)} = N/V$ and local fluctuations around it
are described by the density contrast $\delta^{(\ii)}$. After centering the field appropriately, it is completely fixed by the $N$-point covariance matrix $\tens{C}$,
which we decompose into two-particle submatrices according to
\begin{eqnarray}
 \tens{C} = C_{jk} \otimes E_{jk} \quad \textrm{where} \quad E_{jk} = \ve_j \otimes \ve_k \quad \textrm{and} \nonumber \\
 C_{jk} = \matrix{cc}{ C_{\delta_j \delta_k} & \vec{C}^{\,\T}_{\delta_j p_k} \\ \vec{C}_{p_j \delta_k} & C_{p_j p_k} } \equiv \matrix{cc}{ \ave{\delta^{(\ii)}(\vec{q}^{\,(\ii)}_j) \, \delta^{(\ii)}(\vec{q}^{\,(\ii)}_k)} & \ave{\delta^{(\ii)}(\vec{q}^{\,(\ii)}_j) \,  \vec{P}^{\,(\ii)}(\vec{q}^{\,(\ii)}_k)}^{\T} \\ \ave{\vec{P}^{\,(\ii)}(\vec{q}^{\,(\ii)}_j) \, \delta^{(\ii)}(\vec{q}^{\,(\ii)}_k)} & \ave{ \vec{P}^{\,(\ii)}(\vec{q}^{\,(\ii)}_j) \otimes \vec{P}^{\,(\ii)}(\vec{q}^{\,(\ii)}_k) } } \;.
\label{eq:covariance_matrix_definition}
\end{eqnarray}
Homogeneity and isotropy require that any $C_{jk}$ may only depend on the distance $|\vqi_{jk}| = |\vqi_j - \vqi_k|$. 
Isotropy furthermore implies that a rest frame of the matter distribution must exist in which $\vec{C}_{\delta_j p_k} = 0$ for all $j=k$ and $C_{p_j p_k}$ is a diagonal matrix with all equal entries.
If this was not the case, there would be a preferred direction in the initial momentum field violating isotropy.
Homogeneity implies that the self-correlation matrix of individual particles must be spatially constant and have the diagonal form
\begin{equation}
 C_{jj} = \matrix{cc}{ C_{\delta_j \delta_j} & \vec{0}^{\,\T} \\ \vec{0} & C_{p_j p_j} } = \matrix{cc}{ \sigma_q^2 & \vec{0}^{\,\T} \\ \vec{0} & \sigma_p^2 \mathcal{I}_d } \;.
\label{eq:single_particle_covariance_matrix}
\end{equation}
As was shown in \cite{Bartelmann2016}, we can apply Poisson sampling to individual field configurations and then average over their Gaussian distribution to obtain the initial phase-space distribution
of our particle ensemble as
\begin{eqnarray}
 \mathcal{P}\left( \txi \right) = \frac{V^{-N}}{\left( (2\pi)^{dN} \det \tens{C}_{pp} \right)^{1/2} } \, \mathcal{C}\left( \frac{\partial}{\ii \partial \tpi} \right) \e^{-\frac{1}{2}\tp^{(\ii)^{\T}} \tens{C}_{pp}^{-1} \, \tpi } \;,
\label{eq:initial_phase_space_distribution}
\end{eqnarray}
where we introduced $\tens{C}_{pp} = C_{p_j p_k} \otimes E_{jk}$. While the particle momentum correlations inherit the Gaussian
nature of the momentum density field, the density and density-momentum correlations are contained in the polynomial expression
\begin{eqnarray}
  \mathcal{C}\left( \frac{\partial}{\ii \partial \tpi} \right) 
  &= \prod_{\{n\}} \left( 1 - \ii \sum_{m=1}^N \vec{C}_{\delta_n p_m} \cdot \frac{\partial}{ \ii \vpi_m }\right) 
   + \sum_{\{i<j\}} C_{\delta_i \delta_j} \prod_{\{n\}'} \left( 1 - \ii \sum_{m=1}^N \vec{C}_{\delta_n p_m} \cdot \frac{\partial}{ \ii \vpi_m }\right) \nonumber \\
  &\phantom{{}={}} + \sum_{\{\{i<j\},\{k<l\}\}_{i<k}'} C_{\delta_i \delta_j} C_{\delta_k \delta_l} \prod_{\{n\}'} \left( 1 - \ii \sum_{m=1}^N \vec{C}_{\delta_n p_m} \cdot \frac{\partial}{ \ii \vpi_m }\right) + \ldots
\label{eq:initial_correlation_polynomial}
\end{eqnarray}
The label set $\{n\}$ contains all particle labels $1,\ldots,N$. The primed set $\{n\}'$ excludes the labels $i,j$ chosen in the preceding sum over ordered pairs.
This also applies to all higher order terms. The prime on the sum over ordered tupels of ordered pairs indicates that no label may be shared between the pairs in any individual term of the sum.
This scheme is repeated up to the terms which have $\lfloor N/2 \rfloor$ factors of $C_{\delta_i \delta_j}$.

\section{Cluster expansion of the generating functional} \label{sec:cluster_expansion}

\subsection{The initial correlation operator} \label{sec:ini_corr_operator}

The next step consists of inserting the explicit form of the initial phase-space distribution \eref{eq:initial_phase_space_distribution} into the free generating functional defined in
\eref{eq:gen_func_definition_explicit}. It will be advantageous to rewrite the result such that the effect of initial correlations becomes easier to understand.
For this we need to split the covariance matrix of momentum auto-correlations as
\begin{eqnarray}
 \tens{C}_{pp} &= \left( \sum_{j = k} + \sum_{j \neq k} \right) C_{p_j p_k} \otimes E_{jk} \eqqcolon \sigma_p^2 \, \mathcal{I}_d \otimes \mathcal{I}_N  + \tens{\bar{C}}_{pp} \;.
\label{eq:Cpp_split}
\end{eqnarray}
We define the uncorrelated Gaussian distribution specified by $\sigma_p^2$ as
\begin{equation}
 P_{\sigma_p^2}\left( \tpi \right) \coloneqq \frac{1}{\left(2 \pi \sigma_p^2\right)^{dN/2}} \, \e^{ -\tpi \cdot \tpi / (2\sigma_p^2) } \;,
\label{eq:momentum_gaussian_uncorrelated}
\end{equation}
and assemble all other correlations into the operator
\begin{equation}
 \CorOp\left(\hat{\tchi}_p(t_\ii)\right) \coloneqq \hat{\mathcal{C}}\left(\vochi_p(t_\ii)\right) \, \e^{-\frac{1}{2} \vochi^{\T}_p(t_\ii) \, \tens{\bar{C}}_{pp} \, \vochi_p(t_\ii)} \;.
\label{eq:total_corr_operator}
\end{equation}
After some simple calculations detailed in \ref{app:gaussian_gen_functional}, we can write the free generating functional using the above definitions as
\begin{eqnarray}
 Z_{\CC,0}[\tJ,\tK] = \usi{\txi} \frac{P_{\sigma_p^2}}{V^N} \, \CorOp \bpint{\tx}{\txi}{} \upi{\tchi} \e^{ \ii \left( \tchi \cdot \left(\partial_t + \tens{F}\right)\tx + \tJ \cdot \tx + \tK \cdot \tchi \right) } \;.
\label{eq:free_gen_func_corr_operator}
\end{eqnarray}
Note that if we turn off all initial correlations then $\CorOp \rightarrow 1$ and we recover the generating functional of a non-interacting gas with
momentum dispersion $\sigma_p^2$ and uniform distribution in space. The total correlation operator $\CorOp$ induces shifts of the
initial phase-space positions of the particles relative to this reference system.
\footnote{Note that we could also describe initial correlations as a shift relative to a uniform system at rest if we had not split off the one-particle momentum variance in \eref{eq:Cpp_split}.}
This is very similar to the standard scheme for the initial setup of correlations in cosmological $N$-body simulations \cite{Sirko2005a}, where the Zeldovich approximation is applied to a regular
grid or uncorrelated glass state.

This shows that while the free dynamics of the particles can be obtained exactly, initial Gaussian correlations introduce another operator
$\CorOp$ which couples the degrees of freedom of different particles, in addition to the interaction operator. Thus, even
in the free regime we will have to perform an expansion in the number of particles being correlated. However, we will later find that statistical homogeneity truncates this expansion
at the number $n_f$ of $\Phi_f$-fields in the cumulant. 

\subsection{Cluster expansion} \label{sec:cluster}

In the case $\CorOp \to 1$, the free generating functional \eref{eq:free_gen_func_corr_operator} can be factorised into one-particle contributions,
\ie all integrations over phase-space coordinates and auxiliary fields of different particles can be separated from one another.
Since the collective fields $\Phi_f$ \eref{eq:phase_space_density} and $\Phi_B$ \eref{eq:response_field} can also be separated into one-particle contributions,
the free evolution of its cumulants can then be obtained by calculating the generating functional of only a single particle. 
Starting from \eref{eq:gen_func_definition_explicit} we apply $\e^{\ii H \cdot \hat{\Phi}}$ to \eref{eq:free_gen_func_corr_operator} and then use the appropriate one-particle version of
\eref{eq:momentum_gaussian_uncorrelated} to define this for any particle $j$ as the trace operation
\begin{eqnarray}
 \Tr_j \coloneqq \int_{\xin_j} \bpint{\vx_j}{\vxi_j}{}  \upi{\vchi_j} \, \frac{ P_{\sigma_p^2}}{V} \,
   \e^{\ii \left( \vchi_j \cdot \left( \partial_t + F_j \right) \vx_j + \vec{J}_j \cdot \vx_j + \vec{K}_j \cdot \vchi_j + H \cdot \Phi^{(1)}_j \right) } \;.
\label{eq:one_particle_trace}
\end{eqnarray}
The free generating functional would just be the product over $N$ of these identical traces.
In the case $\CorOp \neq 1$, the correlation operator couples particles so that integrals over their respective coordinates can no longer be separated,
\begin{equation}
 Z_{\CC,0}[H,\tJ,\tK] = \left( \prod_{j=1}^N \Tr_j \right) \, \CorOp\left( \tqi, \hat{\tchi}_p(t_\ii) \right) \;,
\label{eq:free_gen_func_trace}
\end{equation}
where the correlation operator still acts to the left on the exponential contained in the trace.
Any factorisation must now be performed in terms of `clusters' which are formed from a subset of all particles due to their pairwise correlations in some term of
$\CorOp$. This concept is essentially the same as the Mayer cluster expansion \cite{Mayer1941}, where the configurational part of the partition sum of an interacting
gas is expanded in the number of particles taking part in interactions.

There exists a very illustrative visualisation of this scheme in terms of Feynman-like diagrams which we adapt from \cite{becker}. A detailed derivation of the form of diagram lines
and the rules constraining their combination into diagrams representing terms of $\hat{C}_{\mathrm{tot}}$ can be found in \ref{app:diagrams_cluster}.
The four fundamental line types representing correlation operators between particle dots $i$ and $j$ are
\begin{eqnarray}
 \mytikz{\hortwopattern[i][j] \ddline{p1}{p2} } &= \hat{C}_{\delta_i \delta_j} \left(\vqi_{ij},\vochi_{p_i}(t_\ii), \vochi_{p_j}(t_\ii) \right) \coloneqq C_{\delta_i \delta_j} \, \e^{-\vochi_{p_i}^{\,\T}(t_\ii) \, C_{p_i p_j} \, \vochi_{p_j}(t_\ii)} \;, \nonumber \\
 \mytikz{\hortwopattern[i][j] \ppline{p1}{p2} } &= \hat{C}_{p_i p_j}\left(\vqi_{ij}, \vochi_{p_i}(t_\ii), \vochi_{p_j}(t_\ii) \right) \coloneqq  \e^{-\vochi_{p_i}^{\,\T}(t_\ii) \, C_{p_i p_j} \, \vochi_{p_j}(t_\ii)} - 1 \;, \nonumber \\
 \mytikz{\hortwopattern[i][j] \dpline{p1}{p2} } &= \hat{C}_{\delta_i p_j}\left(\vqi_{ij}, \vochi_{p_i}(t_\ii), \vochi_{p_j}(t_\ii) \right) \coloneqq \left( -\ii \vec{C}_{\delta_i p_j} \cdot \vochi_{p_j}(t_\ii) \right) \, \e^{-\vochi_{p_i}^{\,\T}(t_\ii) \, C_{p_i p_j} \, \vochi_{p_j}(t_\ii)} \;, \nonumber \\
 \mytikz{\hortwopattern[i][j] \ddlline{p1}{p2} } &= \hat{C}_{(\delta p)^2_{ij}}\left(\vqi_{ij}, \vochi_{p_i}(t_\ii), \vochi_{p_j}(t_\ii) \right) \nonumber \\
                                                 &\coloneqq \left( -\ii \vec{C}_{\delta_i p_j} \cdot \vochi_{p_j}(t_\ii) \right) \, \left( -\ii \vec{C}_{\delta_j p_i} \cdot \vochi_{p_i}(t_\ii) \right) \, \e^{-\vochi_{p_i}^{\,\T}(t_\ii) \, C_{p_i p_j} \, \vochi_{p_j}(t_\ii)} \;.  
 \label{eq:corr_line_types}
\end{eqnarray}
All these lines depend on the relative coordinate distance $\vqi_{ij}$ through the correlation functions.
The diagrams which can be constructed from these four line types are constrained by three rules, which are derived from the form of the
initial phase-space distribution \eref{eq:initial_phase_space_distribution}. 
\begin{description}
 \item[No self-correlation rule:] Subdiagrams of the form \mytikz[-0.1ex]{ \coordinate[pdot] (p) at (0,0); \coordinate (center) at (0,0.2\gs); \draw[gline] circle[radius=0.2\gs, at=(center)];}
                             for any kind of correlation line are \textit{forbidden}.

 \item[No 2-particle loop rule:] Any pair of particles can be connected directly by at most one line.
                This means that the subdiagram $\mytikz{ \hortwopattern \ggline{p1}{p2}[0.3cm] \ggline{p1}{p2}[-0.3cm]}$ for any combination of correlation lines is \emph{forbidden}.
\end{description}
Note that this does not exclude loops involving three particles or more. The third rule is the
\begin{description}
 \item[$\delta$-Line Rule:] No particle may have more than one solid $\delta$-line attached to it, thus all subdiagrams which include
                            $\mytikz[0.5ex]{\coordinate[pdot] (p) at (0,0); \ddline{p}{0.75\gs,0} \ddline{p}{0.5\gs,0.5\gs} }$ are \emph{forbidden}.
\end{description}
The correlation operator \eref{eq:total_corr_operator} can then be obtained by summing over all ordered $n$-tuples
of particle labels, where $2 \leq n \leq N$, and for each of them summing all diagrams one can draw under the above rules into a fixed graphical arrangement
of particle labels representing their ordering.We write this in a schematic way as
\begin{eqnarray}
\CorOp = 1 + \sum_{n=2}^N \sum_{\{j_1 < \ldots < j_n\}} \Diag{n}(j_1,\ldots,j_n) \;.
\label{eq:Ctot_diag_sum}
\end{eqnarray}
To illustrate this, consider the contribution from three particle diagrams representing pure momentum correlations, which according to the above rules are
\begin{eqnarray}
 \Diag{3}(i,j,k)\Big|_{pp} = \mytikz[2.0ex]{\threepattern[i][j][k] \ppline{p1}{p3} \ppline{p2}{p3} } +
                          \mytikz[2.0ex]{\threepattern[i][j][k] \ppline{p1}{p2} \ppline{p1}{p3} } +
                          \mytikz[2.0ex]{\threepattern[i][j][k] \ppline{p2}{p1} \ppline{p2}{p3} } +
                          \mytikz[2.0ex]{\threepattern[i][j][k] \ppline{p1}{p2} \ppline{p2}{p3} \ppline{p3}{p1} }               
 \label{eq:Ctot_pp_three_particles}
\end{eqnarray}

We call any connected subdiagram of $\ell$ particles an $\ell$-\textit{cluster}, \ie those subdiagrams through which one can trace a continuous path. Starting at the four-particle level,
disconnected diagrams will appear in $\CorOp$. Factorising their contribution to the generating functional into connected diagrams, \ie clusters, is the central idea of the 
cluster expansion. Consider for example
\begin{equation}
 \mytikz[2.0ex]{\fourpattern[1][3][5][6] \ppline{p1}{p2} \ppline{p2}{p4} \ppline{p4}{p3}} \times \mytikz[2.0ex]{\threepattern[2][4][7] \ddline{p1}{p3} \dpline{p2}{p3} } \times \mytikz[2.0ex]{\vertwopattern[8][10] \dpline{p1}{p2}} \times \mytikz{ \pdot[9] }  \times \mytikz[2.0ex]{\vertwopattern[11][12] \ppline{p1}{p2}} \times \mytikz{ \pdot[13] } \times \ldots \times \mytikz{\pdot[N]} \;.
\label{eq:disconnected_diagram_example}
\end{equation}
For any diagram one can count the number $m_\ell$ of clusters of size $\ell$, where in this case we find $m_1 = N-11, m_2 = 2, m_3 = 1, m_4 = 1$ and zero for all larger clusters.
One collects all diagrams with the same set $\{m_\ell\}$ of numbers into a \textit{cluster configuration}.
Since there are only $N$ particles any $\{m_\ell\}$ must obey the constraint
\begin{equation}
\sum_{\ell=1}^N \, \ell \cdot m_\ell = N \;.
\label{eq:cluster_configuration_constraint}
\end{equation}
As we show in \ref{app:diagrams_cluster}, the sum over all diagrams in a cluster configuration factorises into
products of the free $\ell$-particle cluster generating functional of \textit{connected correlations}
\begin{equation}
 W_0^{\,(\ell)} \coloneqq \frac{1}{\ell!} \, \left( \prod_{j=1}^\ell \Tr_j \right) \; \CDiag{\ell}(1,\ldots,\ell) \;. 
\label{eq:l_cluster_gen_func}
\end{equation}
Here $\CDiag{\ell}(1,\ldots,\ell)$ is the sum of all connected $\ell$-particle diagrams obtained with the same logic as $\Diag{n}$. Due to the integrations contained
in the trace operator this quantity has the same value for any set of $\ell$ particles chosen from the overall $N$ particles, thus we define it in terms of a \textit{representative}
set of particles labeled $1, \ldots, \ell$. We then obtain the free generating functional by summing over all possible cluster configurations
\begin{eqnarray}
  Z_{\CC,0}[H] = N! \sum_{\{m_\ell\}^\ast} \prod_{\ell=1}^N \frac{ \left(\cgf{\ell}[H]\right)^{m_\ell}} {m_\ell!} \;,
\label{eq:free_gen_func_factorised}
\end{eqnarray}
where the asterisk in the sum represents the constraint \eref{eq:cluster_configuration_constraint}.

\subsection{Grand canonical generating functional} \label{sec:grand_canonical_gen_func}

In order to write down a grand canonical generating functional for our system we need a probability distribution $\mathcal{P}(N)$ for the number of particles,
which replaces the constraint on the particle number $N$. Finding $\mathcal{P}(N)$ for a general non-equilibrium system is a formidable
task because it may in principle depend on the complete instantaneous microscopic phase-space configuration $\tx(t)$.

However, we are considering a statistically homogeneous and isotropic system. In the absence of external forces and if the interaction potential also respects
these symmetries, they are conserved in time. If we thus can derive $\mathcal{P}(N)$ at the initial time $t_\ii$ by employing only these two symmetries,
it must also be valid at any later time.

Consider the usual setup where we embed our grand canonical system with volume $V_{\GC}$ 
into a much larger canonical system with particle number $N_{\CC}$ and volume $V_{\CC}$. Due to statistical homogeneity all choices for the location of
the embedded subvolume $V_{\GC}$ are equivalent. Homogeneity also implies that the probability for one of the $N_{\CC}$ particles to be in the subvolume $V_{\GC}$
must be given by $p = V_{\GC} / V_{\CC}$. The number of particles inside the subvolume is then binomially distributed, \ie $N_{\GC} \sim \mathcal{B}(N_{\CC},p)$.
Taking the limit $N_{\CC}, V_{\CC} \rightarrow \infty$ with the mean particle density $\mpd \coloneqq N_{\CC} / V_{\CC} = \mpd^{(\ii)} = \mathrm{const.}$ and
$p \rightarrow 0$, the binomial distribution of $N_{\GC}$ turns into a Poisson distribution. Concentrating on the 
grand canonical subsystem and dropping all suffixes, \ie $V_{\GC} = V$ and $N_{\GC} = N$, we thus find 
\begin{equation}
 \mathcal{P}(N) = \frac{\ave{N}^N}{N!} \, \e^{-\ave{N}} = \frac{\mpd^N V^N}{N!} \, \e^{-\ave{N}} \;.
\label{eq:particle_number_poisson_prob}
\end{equation}

The grand canonical generating functional is now obtained by averaging the canonical one with \eref{eq:particle_number_poisson_prob}. Using \eref{eq:free_gen_func_trace} and 
that the interaction operator in \eref{eq:gen_func_definition_explicit} is independent of particle number, we find
\begin{eqnarray}
 Z_{\GC}[H] &= \sum_{N=0}^\infty \, \mathcal{P}\left(N\right) \, \e^{i \hat{H} \cdot \sigma \cdot \hat{H}} Z^{(N)}_{\CC,0}[H] 
                    = \e^{i \hat{H} \cdot \sigma \cdot \hat{H}} \sum_{N=0}^\infty \frac{\e^{-\ave{N}}}{N!} \left( \prod_{j=1}^N \mpd \, V \Tr_j \right) \hat{C}_{\mathrm{tot}} \nonumber \\
                    &= \e^{-\ave{N}} \e^{i \hat{H} \cdot \sigma \cdot \hat{H}} \sum_{N=0}^\infty \, \frac{1}{N!} \, \left( \prod_{j=1}^N \tilde{\Tr}_j \right) \hat{C}_{\mathrm{tot}} \;.
\label{eq:gc_gen_func_definition}
\end{eqnarray}
The modified trace operator $\tilde{\Tr}$ amounts to replacing $1/V \rightarrow \mpd$ in \eref{eq:one_particle_trace}. From here on we will always understand the modified
trace to be used and drop the tilde. The prefactor $\e^{-\ave{N}}$ can either be absorbed into the normalisation of the path integrals or dropped altogether since we only consider
functional derivatives of $\ln Z_{\GC}$ later on. Using \eref{eq:free_gen_func_factorised} and according to \cite{becker} we can execute the sum over particle numbers
to obtain the free grand canonical generating functional as
\begin{eqnarray}
 Z_{\GC,0}[H] &= \sum_{N=0}^{\infty} \, \frac{1}{N!} \, \sum_{\{m_\ell\}^{\ast}} \, N! \, \prod_{\ell=1}^N \, \frac{\left(\cgf{\ell}[H]\right)^{m_\ell}}{m_\ell !} 
                       = \sum_{\{m_\ell\}} \, \prod_{\ell} \, \frac{\left(\cgf{\ell}[H]\right)^{m_\ell}}{m_\ell !} \nonumber \\
                      &= \prod_{\ell=1}^{\infty} \, \sum_{m_\ell=0}^{\infty} \, \frac{\left(\cgf{\ell}[H]\right)^{m_\ell}}{m_\ell !} 
                       = \prod_{\ell=1}^{\infty} \, \e^{\cgf{\ell}[H]} = \e^{\sum_{\ell=1}^{\infty} \, \cgf{\ell}[H]} \;.
\label{eq:free_gc_gen_func_definition}
\end{eqnarray}
In the first step we use that first summing over all cluster configurations with the $N$-particle constraint $\{m_\ell\}^{\ast}$ and then summing over all
particle numbers is identical to summing over all cluster configurations without this constraint. In the second line we reorder the sum in terms of cluster size
rather than cluster configurations and then use the series definition of the exponential function.

Readers familiar with quantum field theory will recognize \eref{eq:free_gc_gen_func_definition} as the usual relation between disconnected and connected Feynman diagrams,
\ie the sum of all disconnected diagrams is given by the exponential of the sum of all connected diagrams. The only difference is that diagrams are now physically interpreted as representing
initial phase-space correlations. We will in fact see shortly that they can be evaluated by applying simple rules very similar to those of traditional Feynman diagrams.

\section{Phase-space density cumulants} \label{sec:density_cumulants}

\subsection{General form of cumulants} \label{sec:general_form_cumulants}

Grand canonical cumulants are defined in the same manner as in \eref{eq:cumulants_definition}. In the non-interacting regime the cumulant generating functional reads
\begin{equation}
 W_{\GC,0}[H] = \ln Z_{\GC,0}[H] = \ln \e^{\sum_{\ell=1}^\infty \cgf{\ell}[H]} = \sum_{\ell=1}^\infty \cgf{\ell}[H] \;.
\label{eq:free_gc_connected_gen_func}
\end{equation}
This simple sum makes calculating free cumulants a lot more comfortable when compared to the canonical ensemble. It also shows that the notion
of connected particle correlation diagrams directly translates into connected correlations of collective fields and that this notion is conserved under free motion.

From the definition of the one-particle trace \eref{eq:one_particle_trace} it is trivial to see that
\begin{eqnarray}
 \Tr^{(\ell)} \coloneqq \prod_{j=1}^\ell \Tr_j = e^{\ii H \cdot \PhiOp{\ell}} \usi{\txi} \mpd^\ell P_{\sigma_p^2}\left(\tpi\right) \e^{\ii \, \tJ \cdot \btx} \;,
\label{eq:l_particle_trace}
\end{eqnarray}
where bold vectors are now defined for the set of $\ell$ representative particles, $\PhiOp{\ell}$ is the corresponding collective field operator and $\btx$ is the free solution
\eref{eq:trajectories_free_solution} for the phase-space trajectories of all particles. A general free cumulant can be written as
\begin{eqnarray}
 \Gf_{\alpha_1(1) \cdots \alpha_n(n)} &= \hat{H}_{\alpha_1(1)} \ldots \hat{H}_{\alpha_n(n)} \left( \sum_{\ell=1}^\infty \, \cgf{\ell}[H,\tens{J},\tens{K}] \right) \, \Big|_{0} \nonumber \\
                                      &= \sum_{\ell=1}^\infty \Tr^{(\ell)} \hat{\Phi}_{\alpha_1}^{(\ell)}(1) \ldots \hat{\Phi}_{\alpha_n}^{(\ell)}(n) \, \frac{1}{\ell!} \CDiag{\ell}(1,\ldots,\ell)\, \Big|_0 \nonumber \\
                                      &\eqqcolon \sum_{\ell=1}^\infty \Gf[\ell]_{\alpha_1(1) \, \ldots \, \alpha_n(n)} \;,
\label{eq:general_cumulant_operator_product}
\end{eqnarray}
where general free $\ell$-particle cumulants were defined in the last line. 
The product of collective field operators in \eref{eq:general_cumulant_operator_product} can be written out into a sum of $\ell^{n}$ terms with the help of \eref{eq:phase_space_density_operator}
and \eref{eq:response_field_operator}, each of them being a product of $n$ single particle operators. In every such term each field label $1,\ldots,n$ appears exactly once.
We will use the notion that some particle $j$ \textit{carries} such a label if the corresponding operator $\hat{\Phi}^{(1)}_{f_j}(1)$ is present.

Statistical homogeneity enforces certain constraints on the two-particle correlations represented by diagram lines. In \ref{app:op_eval_hom_restriction} we show how these lead to the
\begin{description}
 \item[Homogeneity Restriction:] \textit{Such terms in the operator product $\PhiOp{\ell}_{\alpha_1}(1) \ldots \PhiOp{\ell}_{\alpha_n}(n)$ vanish in which at least one of the particles
                                         $j \in \{1,\ldots,\ell\}$ carries no field label.}
\end{description}
The physical picture behind this is best understood by considering the simple example of the two-particle contribution to the two-point $\Phi_f$ cumulant.
If we ask for the Fourier mode contribution of the first particle $a$ to both $\Phi_f$ fields, \ie the particle carries both field labels,
the second particle $b$ can only influence the result due to its initial correlation with $a$. But since we also need to integrate the initial position of $b$ over all of space,
these correlations must cancel out in a statistically homogeneous situation, leading to a factor of zero.

If we only consider pure phase-space density cumulants $\Gf_{f(1) \cdots f(n)}$ this has the important consequence that
\begin{equation}
  \Gf[\ell]_{f(1) \cdots f(n)} = 0 \quad \mathrm{if} \quad n < \ell \quad \Rightarrow \quad \Gf_{f(1) \cdots f(n)} = \sum_{\ell=1}^n \Gf[\ell]_{f(1) \, \ldots \, f(n)} \;,
\label{eq:particle_sum_truncation}
\end{equation}
because with each field label appearing exactly once in every term of the operator product, each term will have at least one particle not carrying a label if $n<\ell$, so they all vanish.
We will extend this statement to mixed cumulants including response fields in section \ref{sec:mixed_cumulants}.

Statistical homogeneity thus naturally truncates the sum over particle numbers in \eref{eq:general_cumulant_operator_product} at the number of phase-density fields $\Phi_f$ present in the cumulant.
This has a very important implication. Since the diagrammatic rules of section \ref{sec:cluster} impose a finite limit on the possible number of diagrams
for a fixed number of particles, we can write down \textit{exact and explicit} expressions accessible to numerical evaluation for any given non-interacting $n$-point cumulant.
In a separate paper, we will argue that one cannot do this in the Eulerian standard perturbation theory of cosmic structure formation.

The general form of an $\ell$-particle phase-space density cumulant $\Gf[\ell]_{f(1) \cdots f(n)}$ is best written down in terms of \textit{field label sets}.
In each non-vanishing individual term of $\PhiOp{\ell}_{\alpha_1}(1) \ldots \PhiOp{\ell}_{\alpha_n}(n)$ in \eref{eq:general_cumulant_operator_product}
the field labels $1,\ldots,n$ are naturally grouped into $\ell$ non-empty sets $\II_j$ being carried by particles $j = 1,\ldots,\ell$. 

We can now change our perspective and only specify \textit{how field labels are grouped} into a collection of sets \textit{without specifying which particle} carries which set.
We call such a collection a \textit{label grouping}. It turns out that summing over all possibilities of assigning sets to particles merely cancels the factor of $1/\ell!$ in
\eref{eq:general_cumulant_operator_product}. For any grouping, we may thus freely choose one arbitrary assignment of label sets to particles by enumerating the collection of sets
as $\II_j$, $j=1,\ldots,\ell$. A general phase-space density cumulant is then simply a sum over all possible groupings 
\begin{eqnarray}
  \Gf[\ell]_{f(1) \cdots f(n)} = \mpd^\ell (2\pi)^d \dirac\left( \sum_{r=1}^n \vec{L}_{\qq,r}(t_\ii) \right) \sum_{\{\II_1,\ldots,\II_\ell\}} e^{-Q(\II_1,\ldots,\II_\ell)} \, \sfcd{\ell}(\II_1,\ldots,\II_\ell) \;.
\label{eq:general_cumulant_sum_groupings}
\end{eqnarray}
The quantity $\sfcd{\ell}$ is defined as the Fourier transformed sum of connected $\ell$-particle correlation diagrams
\begin{equation}
 \sfcd{\ell}(\II_1,\ldots,\II_\ell) \coloneqq \prod_{j=1}^{\ell-1} \int_{\qin_{j\ell}} \e^{-\ii \Lq{j}(t_\ii) \cdot \vqi_{j\ell}} \CDiag{\ell}\left(\II_1, \ldots, \II_\ell\right) \;.
 \label{eq:fourier_sum_con_diagrams}
\end{equation}
The uncorrelated Gaussian momentum distribution $P_{\sigma_p^2}(\tens{p}^{(\ii)})$ in \eref{eq:l_particle_trace} has turned into a Gaussian damping factor
\begin{eqnarray}
\e^{-Q(\II_1,\ldots,\II_\ell)} &\coloneqq  \nexp{ -\frac{\sigma_p^2}{2} \sum_{j=1}^\ell \Lp{j}^{\,2}(t_\ii) } \;,
\label{eq:damping_factor}
\end{eqnarray}
which has been discussed at length in \cite{Bartelmann2017,Dombrowski2017}. The effects of applying collective field operators are parametrised in terms of \textit{phase shift vectors} 
\begin{eqnarray}
 \Lq{j}(t) = \sum_{r \, \in \, \II_j} \vec{L}_{\qq,r}(t) \coloneqq \sum_{r \, \in \, \II_j} \left( \vk_r \, \gqq(t_r,t) + \vl_r \, \gpq(t_r,t) \right) \;, \label{eq:q_shift_vector} \\
 \Lp{j}(t) = \sum_{r \, \in \, \II_j} \vec{L}_{\pp,r}(t) \coloneqq \sum_{r \, \in \, \II_j} \left( \vk_r \, \gqp(t_r,t) + \vl_r \, \gpp(t_r,t) \right) \;. \label{eq:p_shift_vector}
\end{eqnarray}
These are the generalisations of the shift vectors of our previous works \cite{Bartelmann2016,Bartelmann2017} to the complete phase-space.
They encode how the phases of the Fourier factors contributing to the phase-space density field labels $r = (\vk_r,\vl_r,t_r)$
change due to the free motion from time $t$ to time $t_r$ of those particles which carry these field labels. Causality, \ie $t_r \geq t$, is encoded in the propagators.

\subsection{Feynman rules} \label{sec:feynman_rules}

The final step is to evaluate the Fourier transform $\sfcd{\ell}$ \eref{eq:fourier_sum_con_diagrams} of the connected correlation diagrams.
This is easily done be following a set of rules that are very similar to Feynman diagram rules in quantum field theory. These rules are derived in 
\ref{app:feynman_rules_derivation}. The basic idea is to extend the Fourier transforms in \eref{eq:fourier_sum_con_diagrams} from coordinates $\vqi_{j\ell}$ centered on the $\ell$-th coordinate to all
relative coordinates $\vqi_{jk}$. The Fourier transformed correlation lines in the diagrams are then defined as 
\begin{eqnarray}
\mytikz{\hortwopattern[j][k] \ppline{p1}{p2}} &= \mathcal{P}_{p_j p_k}(\Ki_{jk}) \coloneqq  \int_{\qin_{jk}} \e^{-\ii \, \Ki_{jk} \cdot \vqi_{jk}} \hat{C}_{p_i p_j}\left(\vqi_{ij}, \Lp{j},\Lp{k}\right) \;, \nonumber \\
\mytikz{\hortwopattern[j][k] \ddline{p1}{p2}} &= \mathcal{P}_{\delta_j \delta_k}(\Ki_{jk}) \coloneqq  \int_{\qin_{jk}} \e^{-\ii \, \Ki_{jk} \cdot \vqi_{jk}} \hat{C}_{\delta_i \delta_j}\left(\vqi_{ij}, \Lp{j},\Lp{k}\right) \;, \nonumber \\
\mytikz{\hortwopattern[j][k] \dpline{p1}{p2}} &= \mathcal{P}_{\delta_j p_k}(\Ki_{jk}) \coloneqq \int_{\qin_{jk}} \e^{-\ii \, \Ki_{jk} \cdot \vqi_{jk}} \hat{C}_{\delta_i p_j}\left(\vqi_{ij}, \Lp{j},\Lp{k}\right) \;, \nonumber \\
\mytikz{\hortwopattern[j][k] \ddlline{p1}{p2}} &= \mathcal{P}_{(\delta p)^2_{jk}}(\Ki_{jk}) \coloneqq \int_{\qin_{jk}} \e^{-\ii \, \Ki_{jk} \cdot \vqi_{jk}} \hat{C}_{(\delta p)^2_{jk}}\left(\vqi_{ij}, \Lp{j},\Lp{k}\right) \;.
\label{eq:fourier_corr_lines}
\end{eqnarray}
where $j<k$ in all cases by definition. Along each line there is thus a flow of a general \textit{Fourier momentum} $\Ki_{jk}$ from smaller to larger labels. The algorithm for evaluating
\eref{eq:fourier_sum_con_diagrams} for a given grouping $\{\II_1,\ldots,\II_\ell\}$ of field labels is then as follows.
\begin{enumerate}
 \item Choose a fixed graphical arrangement of $\ell$ dots labeled $j=1,\ldots,\ell$ representing the particles carrying the label sets $\II_j$. These are the vertices of the diagrams.
 \item With this fixed set of vertices, draw all possible diagrams using the lines in \eref{eq:fourier_corr_lines} while obeying the rules detailed in \ref{sec:cluster}.
       Repeat the next three steps for all diagrams.
 \item For any closed loop in a diagram, assign one of the lines in the loop connecting vertices labeled $j<k$ a loop momentum $\vki_{jk}$ and introduce an integral $\int_{k^{(\ii)}_{jk}}$.
 \item Pick a vertex $j$ which has only one line left with undetermined momentum and use the conservation law
       \begin{equation}
        \Lq{j} + \sum_{i=1}^{j-1} \Ki_{ij} - \sum_{i=j+1}^{\ell} \Ki_{ji} = 0 
        \label{eq:fourier_momentum_conservation}
       \end{equation}
       to fix the momentum of this line. Incoming momenta from vertices with smaller labels are counted positive while outgoing momenta to vertices with larger labels are counted negative.
       Momenta associated with vertices not connected to vertex $j$ are zero.
 \item Consecutively go through all other vertices to fix the remaining undetermined momenta by repeatedly applying the previous step.
\end{enumerate}

One may always use that the leading Dirac delta distribution in \eref{eq:general_cumulant_sum_groupings} ensures the global conservation of the position phase-shift vectors
$\Lq{1} + \ldots + \Lq{\ell} = 0$. Loop momenta $\vki_{jk}$ may be shifted freely in position Fourier space and general momenta $\Ki_{jk}$ can be reversed in sign because due to statistical
isotropy the correlation functions only depend on the absolute value $|\vqi_{jk}|$ of the separation vectors.

These rules are best understood by considering some easy examples. The most trivial are two-particle cumulants. A general grouping of the field labels $1,\ldots,n$ is given by $\II_1,\II_2$.
The sum of all possible two-particle diagrams is
\begin{eqnarray}
 \mytikz{\hortwopattern[1][2] \ddline{p1}{p2} } +
 \mytikz{\hortwopattern[1][2] \dpline{p2}{p1} } +
 \mytikz{\hortwopattern[1][2] \dpline{p1}{p2} } + 
 \mytikz{\hortwopattern[1][2] \ppline{p1}{p2} } +
 \mytikz{\hortwopattern[1][2] \ddlline{p1}{p2} } \;. 
\label{eq:two_particle_corr_diagrams}
\end{eqnarray}
Since there are no loops we directly go to step (iv) of our algorithm and pick vertex 1. According to our convention, we have to count the general momentum $\Ki_{12}$
as outgoing leading to the conservation law
\begin{equation}
 \Lq{1} - \Ki_{12} = 0 \quad \Rightarrow \quad \Ki_{12} = \Lq{1} \;.
\label{eq:two_particle_mom_conservation}
\end{equation}
We thus find that the sum of correlation diagrams evaluates to
\begin{equation}
 \sfcd{2}(\II_1,\II_2) = \left(\mathcal{P}_{\delta_1 \delta_2}\ + \mathcal{P}_{p_1 \delta_2} + \mathcal{P}_{\delta_1 p_2} + \mathcal{P}_{p_1 p_2} + \mathcal{P}_{(\delta p)^2_{12}}\right)\left(\Lq{1}\right)
\label{eq:two_particle_corr_diag_sum}
\end{equation}
According to \eref{eq:general_cumulant_sum_groupings} the two-particle contribution to the $n$-point phase-space density cumulant is then given by
\begin{equation}
  \Gf[\ell]_{f(1) \ldots f(n)} \coloneqq \mpd^2 (2\pi)^d \dirac\left( \sum_{r=1}^n \vec{L}_{\qq,r}(t_\ii) \right) \sum_{\{\II_1,\II_2\}} \e^{-\frac{\sigma_p^2}{2} \left(\Lp{1}^{\,2}(t_\ii) + \Lp{2}^{\,2}(t_\ii)\right)} \, \sfcd{2}(\II_1,\II_2) \;.
\label{eq:general_two_particle_cumulant}
\end{equation}
In order to demonstrate the determination of Fourier momenta when loops are involved we consider a simple one-loop diagram of four particles.
\begin{equation}
\fl \mytikz[4.0ex]{\largefourpattern[1][2][3][4] 
        \draw[pline] (p1) -- node[below,sloped,scale=0.65]{$\Ki_{12}$} (p2);
        \draw[pline] (p1) -- node[above,sloped,scale=0.65]{$\vki_{13}$} (p3);
        \draw[dline] (p2) -- ($(p2)!0.45!(p3)$);
        \draw[pline] ($(p2)!0.45!(p3)$) -- (p3);
        \path (p2) -- node[below,sloped,scale=0.65]{$\Ki_{23}$} (p3);
        \draw[dline] (p3) -- node[above,sloped,scale=0.65]{$\Ki_{34}$} (p4);
        }
 \rightarrow
 \mytikz[4.0ex]{\largefourpattern[1][2][3][4] 
        \draw[pline] (p1) -- node[below,sloped,scale=0.65]{$\Lq{1} - \vki_{13}$} (p2);
        \draw[pline] (p1) -- node[above,sloped,scale=0.65]{$\vki_{13}$} (p3);
        \draw[dline] (p2) -- ($(p2)!0.45!(p3)$);
        \draw[pline] ($(p2)!0.45!(p3)$) -- (p3);
        \path (p2) -- node[below,sloped,scale=0.65]{$\Ki_{23}$} (p3);
        \draw[dline] (p3) -- node[above,sloped,scale=0.65]{$\Ki_{34}$} (p4);
        }  
  \rightarrow
 \mytikz[4.0ex]{\largefourpattern[1][2][3][4] 
        \draw[pline] (p1) -- node[below,sloped,scale=0.65]{$\Lq{1} - \vki_{13}$} (p2);
        \draw[pline] (p1) -- node[above,sloped,scale=0.65]{$\vki_{13}$} (p3);
        \draw[dline] (p2) -- ($(p2)!0.45!(p3)$);
        \draw[pline] ($(p2)!0.45!(p3)$) -- (p3);
        \path (p2) -- node[below,sloped,scale=0.65]{$\Lq{1} + \Lq{2} - \vki_{13}$} (p3);
        \draw[dline] (p3) -- node[above,sloped,scale=0.65]{$\Ki_{34}$} (p4);
        }  
  \rightarrow
   \mytikz[4.0ex]{\largefourpattern[1][2][3][4] 
        \draw[pline] (p1) -- node[below,sloped,scale=0.6]{$\Lq{1} - \vki_{13}$} (p2);
        \draw[pline] (p1) -- node[above,sloped,scale=0.6]{$\vki_{13}$} (p3);
        \draw[dline] (p2) -- ($(p2)!0.45!(p3)$);
        \draw[pline] ($(p2)!0.45!(p3)$) -- (p3);
        \path (p2) -- node[below,sloped,scale=0.6]{$\Lq{1} + \Lq{2} - \vki_{13}$} (p3);
        \draw[dline] (p3) -- node[above,sloped,scale=0.6]{$\Lq{1} + \Lq{2} + \Lq{3}$} (p4);
        }  
\label{eq:one_loop_diagram_example}
\end{equation}
In the first diagram we have chosen to assign the loop momentum to the line connecting vertices $1$ and $3$. We then go through the vertices $1$, $2$ and $3$ consecutively and for each of them fix the
undetermined $\Ki_{ij}$ by using the conservation law \eref{eq:fourier_momentum_conservation}.
\begin{eqnarray}
 \Lq{1} - \Ki_{12} - \vki_{13} = 0 \quad \Rightarrow \quad \Ki_{12} = \Lq{1} - \vki_{13} \;, \nonumber \\
 \Lq{2} + \Ki_{12} - \Ki_{23} = 0 \quad \Rightarrow \quad \Ki_{23} = \Lq{1} + \Lq{2} - \vki_{13} \;, \nonumber \\
 \Lq{3} + \Ki_{23} + \vki_{13} - \Ki_{34} = 0 \quad \Rightarrow \quad \Ki_{34} = \Lq{1} + \Lq{2} + \Lq{3} \;.
\label{eq:one_loop_example_fix_K12}
\end{eqnarray}
This particular contribution to $\Sigma^{(\ell)}_{\mathrm{con}}(\II_1,\ldots,\II_4)$ would then be 
\begin{equation}
\fl \mathcal{P}_{\delta_3 \delta_4}\left(\Lq{1}+\Lq{2}+\Lq{3}\right) \int_{\kin_{13}} \mathcal{P}_{p_1 p_2}\left(\Lq{1}-\vki_{13}\right) \mathcal{P}_{\delta_2 p_3}\left(\Lq{1}+\Lq{2}-\vki_{13}\right) \mathcal{P}_{p_1 p_3}\left(\vki_{13}\right) \;. 
\label{eq:}
\end{equation}

\subsection{Derived cumulants} \label{sec:derived_cumulants}

Preserving the full microscopic phase-space statistics by working in terms of the phase-space density $\Phi_f$ has the benefit that at any $n$-point level, once the corresponding
$\Phi_f$-cumulant \eref{eq:general_cumulant_sum_groupings} has been calculated, one can directly derive cumulants between any observables that can be defined in terms of $\Phi_f$.
Momentum moments are of special interest since they lead to particle density and velocity density statistics. These are defined as integrals over real momentum space
whereas the cumulants \eref{eq:general_cumulant_sum_groupings} are calculated in Fourier momentum space. We thus need to transform the Fourier momentum of the external
label under consideration back into real momentum, multiply by the appropriate factors of momentum and integrate over momentum space. For some field label $r$ this leads to the operator 
\begin{equation}
  \int_{p_r} p_r^{i_1} \ldots p_r^{i_m} \int_{l_r} \e^{\ii \, \vl_r \cdot \vp_r} \, \ldots = (-1)^m \int_{l_r} \, \dirac\left(\vl_r\,\right) \left( \frac{\partial}{\ii \partial l_r^{i_1}} \ldots \frac{\partial}{\ii \partial l_r^{i_m}} \right) \dots \;,
\label{eq:momentum_moment_operator}
\end{equation}
for the $m$-th order moment, where $i_a \in \{1,\ldots,d\} \; \forall a \in \{1,\ldots,m\}$. The $n$-point particle density cumulant is particularly straightforward
since it is the zeroth order moment in all $\Phi_f$ entries,
\begin{equation}
 \Gf[\ell]_{\rho(1) \ldots \rho(n)} = \int_{l_1} \ldots \int_{l_n} \dirac\left(\vl_1\right) \ldots \dirac\left(\vl_n\right) \Gf[\ell]_{f(1) \ldots f(n)} \;.
\label{eq:n_point_density_cumulant}
\end{equation}

As an explicit example, let us consider the two-point particle density cumulant $\Gf_{\rho(1) \rho(2)}$ for a 3D-system of particles on straight trajectories with $\gqq = 1$ and $\gpq = 0$.
We set $\vl_1 = \vl_2 = 0$ according to \eref{eq:n_point_density_cumulant}. This reduces the phase-shift vectors \eref{eq:q_shift_vector} and \eref{eq:p_shift_vector}
to $\vec{L}_{\qq,r}(t_\ii) = \vec{k}_r$ and $\vec{L}_{\pp,r}(t_\ii) = \gqp(t_r, t_\ii) \, \vec{k}_r$. According to the homogeneity restriction the only non shot-noise
contribution is the two-particle term. There is only one possible grouping of the field labels $1,2$ into two non-empty sets. We choose to enumerate them as $\II_1 = \{1\}, \II_2 = \{2\}$.
Setting $t_\ii = 0$ and $t_1=t_2=t$, we use the general result \eref{eq:general_two_particle_cumulant} to find
\begin{eqnarray}
 \fl G^{(0,2)}_{\rho(1) \rho(2)} &= \mpd^2 (2\pi)^3 \dirac\left(\vk_1 + \vk_2\right) e^{-\sigma_p^2 \, k_1^2 \, \gqp^2(t,0)}
 \int_{\qin_{12}} e^{-\ii\vk_1 \cdot \vqi_{12}} \nonumber \\
 \fl &\phantom{{}={}} \left[ \left( 1 + C_{\delta_1 \delta_2} - 2\ii \gqp(t,0) \vec{C}_{\delta_1 p_2} \cdot \vk_1 - \left( \gqp(t,0) \vec{C}_{\delta_1 p_2} \cdot \vk_1 \right)^2 \right) e^{\gqp^2(t,0) \, \vk^{\,\T}_1 C_{p_1 p_2} \vk_1} - 1 \right] \;.
\label{eq:two_particle_density_cumulant}
\end{eqnarray}
We want to emphasize at this point that besides the assumption of statistical homogeneity no approximation or perturbative expansion has gone into deriving this explicit expression.
The underlying microscopic physics driving the time evolution of the initially Gaussian density field encompasses the entirety of its free streaming kinematics including all effects
that must be considered non-linear at the level of macroscopic fields. Since velocity density statistics will be derived from the same fundamental $\Phi_f$-cumulants this feature will apply to them as well.

\section{Mixed cumulants and causality} \label{sec:mixed_cumulants}

So far we only considered the explicit calculation of $\Phi_f$-cumulants. However, when calculating perturbative corrections in the particle interaction
we also need to use mixed cumulants of both $\Phi_f$ and the response field $\Phi_B$.
According to \eref{eq:response_field_operator} the single particle $\PhiOp{1}_{B_j}$-operator is the product of the response factor operator $\hat{b}_j$
and the phase-space density operator $\PhiOp{1}_{f_j}$. In any term of the operator product $\PhiOp{\ell}_{\alpha_1}(1) \ldots \PhiOp{\ell}_{\alpha_n}(n)$ of
\eref{eq:general_cumulant_operator_product}, we may thus push all density operators to the right and execute them first.

The $\hat{b}_j$ contain only derivatives \wrt to the $\vec{K}_{p_j}$ sources which do not couple to the
initial phase-space coordinates and may be pulled in front of the integral over them. Mixed cumulants can thus always be derived from pure $\Phi_f$ cumulants  
by multiplying evaluated terms in the operator product with the appropriate response factors $b_j$ resulting from application of operators $\hat{b}_j$.
By combining \eref{eq:mom_shift_operator_evaluation} with \eref{eq:response_field_operator} we find these response factors to be given by
\begin{eqnarray}
 b_j(u) = \ii \, \Lp{j}(t_u) \cdot \vec{k}_u\;.
\label{eq:response_factor} 
\end{eqnarray}
The physical interpretation is straightforward. The spatial gradient in \eref{eq:response_field_operator} has been Fourier transformed into $\vk_u$.
Multiplication by a potential $v(k_u)$ and integration over $\vk_u$ will generate a force onto particle $j$ leading to an instantaneous momentum change $\delta \vp$ at time $t_u$.
The corresponding response of the Fourier phase contribution of particle $j$ to the phase-space density at the field labels in $\II_j$ is then evolved under
free motion forward in time with the particle propagator $\mathcal{G}$ contained in the momentum phase shift vector.

In order to write down a general mixed cumulant we denote the response factors as $b_{\II(r)} \coloneqq b_j(r)$, with $\II_j = \II(r)$ the set containing the field label $r$.
A \textit{mixed cumulant} is then given by
\begin{eqnarray}
 \Gf[\ell]_{f(1) \cdots f(n_f) B(1') \cdots B(n'_B)} &= \mpd^\ell (2\pi)^d \dirac\left( \sum_{r=1}^n \vec{L}_{\qq,r}(t_\ii) \right) \nonumber \\
 &\phantom{{}={}}\sum_{\{\II_1, \ldots, \II_\ell\}} e^{-Q(\II_1, \ldots, \II_\ell)} \left(b_{\II(1')} \ldots b_{\II(n'_B)}\right) \sfcd{\ell}(\II_1,\ldots,\II_\ell)  \,.
\label{eq:general_mixed_cumulant}
\end{eqnarray}

We now take a closer look at the causal structure of these response factors. We first recognise that in any perturbative
calculation aiming at the statistics of physical observables, \ie cumulants of $\Phi_f$ and integrals thereof, every $\Phi_B$-field will always appear
in combination with the $\sigma$-matrix \eref{eq:interaction_matrix} of particle interactions. For such calculations it is thus advantageous to introduce a new collective
field directly in Fourier space as
\begin{equation}
 \Phi_\mathcal{F}(r) \coloneqq \usi{\bar{r}} \sigma_{fB}(r,-\bar{r}) \, \Phi_B(\bar{r}) \;.
\label{eq:force_field_definition}
\end{equation}
Its physical content is just what we described following \eref{eq:response_factor}, \ie it gives the linear response of the system per phase-space density $\Phi_f$ at the Fourier phase-space point $r$
and as such must always be contracted with another $\Phi_f$ field appearing in the perturbation series, \eg $\Gf[1]_{f(1) \mathcal{F}(\bar{1})} \cdot \Gf[2]_{f(-\bar{1}) f(2)}$.
Since the $\sigma$-matrix only depends on field labels, it does not interfere with any of our previous calculations and we can obtain cumulants involving $\Phi_\mathcal{F}$ as \textit{compound cumulants}
by contracting mixed cumulants with the $\sigma$-matrix,
\begin{eqnarray}
 \Gf[\ell]_{f(1) \cdots f(n_f) \, \mathcal{F}(1') \cdots \mathcal{F}(n'_B)} \coloneqq \prod_{r=1}^{n_B} \left( \usi{\bar{r}} \sigma_{fB}(r', -\bar{r}) \right) \Gf[\ell]_{f(1) \cdots f(n_f) B(\bar{1}) \cdots B(\bar{n}_B)} \;. 
\label{eq:compound_cumulant}
\end{eqnarray}

Since we assume particle interactions to act instantaneously we have $\sigma_{fB}(r',-\bar{r}) \propto \dirac(t_{r'}-t_{\bar{r}})$, so we can always set $t_{r'} = t_{\bar{r}}$ from here on
and consider $\Phi_B$-labels and $\Phi_\mathcal{F}$-labels equivalent for causal considerations.
Since we also describe the interaction by a purely space-dependent potential we obtain the factor $\dirac(\vec{l}_r)\,\dirac(\vec{l}_{r'})$
in the $\sigma(r,r')$-matrix \eref{eq:interaction_matrix}. If we combine this with $\gqp(t,t) = 0$, we see that for compound cumulants \eref{eq:compound_cumulant} we only ever
need to consider effective $b_j$ factors without instantaneous response,
\begin{equation}
 b_j(\bar{u}) = \ii \sum_{r \, \in \, \{\II_j \smallsetminus \bar{u}\}} \vec{L}_{\pp,r}(t_{\bar{u}}) \cdot \vk_{\bar{u}} \;.
\label{eq:response_factor_wo_inst_response}
\end{equation}

Let us generalise our arguments following \eref{eq:response_factor}. We can imagine the individual phase-shift factors as they appear
in \eref{eq:response_factor_wo_inst_response} as arrows connecting dots representing points in time, where the direction of the arrow is given by the time ordering of the propagator.
For a compound cumulant \eref{eq:compound_cumulant}, let the subset $\{\bar{u}_1,\ldots,\bar{u}_m\} \subset \{\bar{1},\ldots,\bar{n}_B\}$ be carried by particle $j$,
\ie $\{\bar{u}_1,\ldots,\bar{u}_m\} \subset \II_j$. The product of response factors $b_j$ is then
\begin{equation}
 b_j(\bar{u}_1) \ldots b_j(\bar{u}_m) \rightarrow \left(\sum_{r \, \in \, \{\II_j \smallsetminus \bar{u}_1\}} \mytikz{\hortwopattern[t_{\bar{u}_1}][t_r] \arrline{p1}{p2} }\right) \ldots
 \left(\sum_{r \, \in \, \{\II_j \smallsetminus \bar{u}_m\}} \mytikz{\hortwopattern[t_{\bar{u}_m}][t_r] \arrline{p1}{p2}}\right) \;.
\label{eq:response_factor_diagrams}
\end{equation}
If we write this out into a sum, each individual term is a `time flow' diagram, where arrows represent how the effects of interactions on particle $j$ at some instance of time `flow' with time
to some other field label. This other field label may in turn also be an instance where the particle interacts and again the effect needs to flow to a field with another label.
The sum of all these diagrams thus represents all possible time orderings of interaction events for particle $j$ given the set $\II_j$ of field labels it carries.

As an example consider a particle $j$ carrying the set $\II_j = \{\bar{u}_1,\ldots,\bar{u}_5, r\}$, \ie carrying five $\Phi_B$-labels and the $\Phi_f$-label $r$
without loss of generality. One possible flow diagram would be
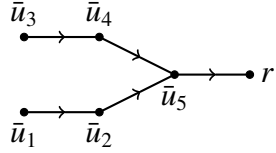
\begin{center}
\tikz[baseline=4.0ex]{
  \path (0,0) coordinate[pdot] (t1)
              ++ (1.0,0.0) coordinate[pdot] (t2)
              ++ (0.0,1.0) coordinate[pdot] (t3)
              ++ (-1.0,0.0) coordinate[pdot] (t4)
              ++ (2.0,-0.5) coordinate[pdot] (t5)
              ++ (1.0,0.0) coordinate[pdot] (t6);  
              \arrline{t1}{t2} \arrline{t4}{t3} \arrline{t2}{t5} \arrline{t3}{t5} \arrline{t5}{t6}
              \node[below] at (t1) {$\bar{u}_1$};
              \node[below] at (t2) {$\bar{u}_2$};
              \node[above] at (t3) {$\bar{u}_4$};
              \node[above] at (t4) {$\bar{u}_3$};
              \node[below] at (t5) {$\bar{u}_5$};
              \node[right] at (t6) {$r$};
 }
\end{center}
and of course all possible permutations of the $\bar{u}_k$. 

The only ab initio topological restriction on these time flow diagrams resulting from their definition \eref{eq:response_factor_diagrams} is that all $\Phi_B$-labels
$\bar{u}_1, \ldots, \bar{u}_m \in \II_j$ appear exactly once as a dot in every diagram, have exactly one line line pointing \emph{away} from them and that any $\Phi_f$-label $r \in \II_j$
may only appear as an end-point of time flow. In \ref{app:time_loops} we show the additional restriction that due to causality there may be \textit{no closed loops} in these diagrams.
If we combine these restrictions we see that all possible diagrams must be composed of tree-like structures with an unambigious direction of time flow terminating at some $\Phi_f$-label $r$
(compare also with the diagrams in \cite{Bartelmann2017}).
Consider again the above example and assume \eg $t_r < t_{\bar{u}_5}$, then the diagram immediately vanishes since
$\mytikz{\path (0,0) coordinate[pdot] (t1) ++ (1.0,0.0) coordinate[pdot] (t2);
         \arrline{t1}{t2}
         \node[left] at (t1) {$\bar{u}_5$};
         \node[right] at (t2) {$r$};} \propto \heavi(t_r - t_{\bar{u}_5})$.
Also, if there is no $\Phi_f$-label, time must flow from the $\bar{u}_5$-dot to some other $\bar{u}_i$-dot which creates a closed loop.

For the compound cumulants these topological restrictions translate into the
\begin{description}
 \item[Causality Restriction:] \textit{Only such field label groupings $\{\II_1,\ldots,\II_\ell\}$ give non-vanishing contributions to compound cumulants \eref{eq:compound_cumulant},
				       which have the following property:
				       For all $\II_j, j=1,\ldots,\ell$ there must be at least one $\Phi_f$-label $r \in \II_j$ such that $t_{r} \geq t_{\bar{u}} = t_{u'}$
				       for all $\Phi_B$-labels $\bar{u} \in \II_j$.}
\end{description}
This has the direct consequence that
\begin{equation}
 \Gf[\ell]_{f(1) \cdots f(n_f) \, \mathcal{F}(1') \cdots \mathcal{F}(n'_B)} = 0 \quad \mathrm{if} \; \exists u' \in 1',\ldots,n'_B \; \mathrm{with} \; t_{u'} > t_{r} \, \forall r \in 1,\ldots,n_f \;.
\label{eq:compound_cumulant_causality}
\end{equation}
Even more importantly, we can combine both the causality and the homogeneity restriction to obtain a truncation criterion for compound cumulants
\begin{eqnarray}
 &\Gf[\ell]_{f(1) \cdots f(n_f) \mathcal{F}(1') \cdots \mathcal{F}(n'_B)} =  0 \quad \textrm{if} \quad n_f < \ell \nonumber \\
 \Rightarrow& \Gf_{f(1) \cdots f(n_f) \mathcal{F}(1') \cdots \mathcal{F}(n'_B)} = \sum_{\ell=1}^{n_f} \Gf[\ell]_{f(1) \cdots f(n_f) \mathcal{F}(1') \cdots \mathcal{F}(n'_B)} \;.
\label{eq:compound_cumulant_truncation}
\end{eqnarray}
Adding any number of $\Phi_B$-labels to the case of \eref{eq:particle_sum_truncation} will still leave particles carrying no labels or only $\Phi_B$-labels, so again all terms vanish.
This shows that not only can we write down exact and explicit cumulants in the non-interacting regime, but also that at any fixed order of perturbation theory in the interaction we can account for all
effects of initial correlations coupling freely streaming particles with a finite number of perturbative terms.

\section{Conclusions} \label{sec:conclusions}

Following up on previous work in \cite{Bartelmann2016,Bartelmann2017} we refined the treatment of non-interacting cumulants of collective fields in the framework of KFT.
We extended the position-space density to the complete phase-space density and included all initial correlations of a Gaussian density and velocity field in an exact fashion.
The latter was achieved by condensing the effects of the complicated initial phase-space distribution \eref{eq:initial_phase_space_distribution} into a coupling operator \eref{eq:total_corr_operator}
between particles acting at the initial time. We saw that the idea of how KFT incorporates initial correlations is actually quite close to the initial
setup scheme of cosmological $N$-body simulations in that it applies phase-space shifts to an uncorrelated system.

We expressed this correlation operator in a diagrammatic language with rules derived from the form of the initial phase-space distribution \eref{eq:initial_phase_space_distribution}.
Using the principles of the Mayer cluster expansion we factorised the free generating functional into connected diagrams. Exploiting the statistical
homogeneity and isotropy of our system we changed our description to a grand canonical ensemble which made the calculation of cumulants of the collective fields straightforward.
We saw that the notion of connected correlations of collective fields is equivalent to connected particle diagrams in the initial correlations and that this notion is conserved under free evolution.

We then re-derived the general scheme of how to calculate such cumulants for the full phase-space density $\Phi_f$. As shown in \ref{app:diagrams_cluster} and \ref{app:feynman_rules_derivation},
the use of a diagrammatic approach allows us to work out all tedious combinatorics in a general way such that the calculation of any cumulant can be reduced to evaluating all possible correlation
diagrams according to familiar Feynman-type rules without the need for any kind of symmetry factors.
We saw that the combination of statistical homogeneity and causal consistency of interaction events impose restrictions on which kind of terms 
may contribute to given free and compound cumulants. These restrictions naturally truncate the number of particles one has to consider at the $n_f$-point order of $\Phi_f$-fields in
the cumulant. At any given order of perturbation theory in the interaction we can thus in principle obtain results for the time evolution of phase-space statistics
which take the full non-linear coupling of free streaming kinematics due to initial correlations into account with only a finite number of explicit terms.

This result also has the profound consequence that in the non-interacting regime KFT generates exact time-evolved cumulants of a Gaussian density field.
We gave the explicit result for the two-point density cumulant in \eref{eq:two_particle_density_cumulant},
\ie the density power spectrum which is of prime interest in cosmology. We will show in a separate paper that this is next to impossible to achieve in the Eulerian approach of SPT,
which includes gravity at the linear level at the price of giving up knowledge of the exact free dynamics. The linear growth behaviour of cosmic density fluctuations derived in linear
SPT is however easily recovered in a resummed form of KFT perturbation theory without losing any information about the free dynamics. This result will be developed in a separate paper, where our insights
into the causal structure of compound cumulants obtained in the present work will prove beneficial for understanding the causal flow through entire terms in perturbative expansions.

We have not presented any numerical evaluation of quantities like \eref{eq:two_particle_density_cumulant} because implementing the necessary Fourier transforms \eref{eq:fourier_corr_lines}
of the correlation lines is a quite formidable task due to very large dynamic range of scales that must be considered and is thus beyond the scope of this work.
For numerical evaluations of the density fluctuation power spectrum restricted to initial momentum correlations we refer the reader to \cite{Bartelmann2017,Dombrowski2017}.
In future work, we will endeavour to extend these calculations to include all correlation types in order to test the free theory against non-interacting $N$-body simulations.

Nonetheless it is obvious that once \eref{eq:fourier_corr_lines} is precomputed for general arguments, the presented approach should be well suited
for automated calculation of arbitrary free cumulants by a computer. The idea is the same as in \cite{Bartelmann2017} in that one specifies the desired cumulant, and a symbolic code generates the appropriate
correlation diagrams which can then be evaluated according to the Feynman rules presented in section \ref{sec:feynman_rules}.
This would then serve as a basis for perturbative and resummed treatments of the interacting system.

Beyond this application to cosmic structure formation our results should be applicable with little modification to any system which is statistically homogeneous and whose dynamics
can be described in terms of an Hamiltonian with an interaction respecting homogeneity. Two examples would be the statistical evolution of both cold Rydberg gases and classical spin lattices.
\hfill \\
\hfill \\
\textbf{Acknowledgements}
We are grateful for many helpful comments and discussions to Carsten Littek, Celia Viermann, Johannes Dombrowski, Sara Konrad and Björn Schäfer.
This work was supported in part by the German Excellence Initiative and by the Collaborative Research Centre TR 33 ``The Dark Universe'' of the German
Science Foundation.

\appendix
\section{Generating functional for Gaussian initial conditions} \label{app:gaussian_gen_functional}

For the following calculations it will be advantageous to define
\begin{eqnarray}
 \tens{\bar{J}}^{\T}_q \coloneqq \bsi{t}{t_\ii}{t_\ff} \tJ(t)^{\T} \tens{\mathcal{G}}(t,t_\ii) \, \tens{\mathcal{P}}_q \quad \textrm{where} \quad \tens{\mathcal{P}}_q = \cvector{ \mathcal{I}_d \\ 0 } \otimes \mathcal{I}_N \;, 
\label{eq:time_averaged_Jq} \\
 \tens{\bar{J}}^{\T}_p \coloneqq \bsi{t}{t_\ii}{t_\ff} \tJ(t)^{\T} \tens{\mathcal{G}}(t,t_\ii) \, \tens{\mathcal{P}}_p \quad \textrm{where} \quad \tens{\mathcal{P}}_p = \cvector{ 0 \\ \mathcal{I}_d } \otimes \mathcal{I}_N \;.
\label{eq:time_averaged_Jp}
\end{eqnarray}
With these we can write the free generating functional \eref{eq:free_gen_func} with \eref{eq:initial_phase_space_distribution} inserted as
\begin{eqnarray}
 Z_{\CC,0}[\tJ,\tK] = \frac{\mathcal{N}_p}{V^N} \usi{\txi} & \e^{\ii \left( \tens{\bar{J}}_q \cdot \tqi + \tens{\bar{J}}_p \cdot \tpi - \tJ \cdot \tens{\mathcal{G}} \cdot \tK \right)} \, \mathcal{C}\left( \frac{\partial}{\ii \partial \tpi} \right) \e^{-\frac{1}{2}\tp^{(\ii)^{\T}} \tens{C}_{pp}^{-1} \, \tpi } \;,
\label{eq:free_gen_func_explicit}
\end{eqnarray}
where $\mathcal{N}_p$ is the normalisation factor of the Gaussian in \eref{eq:initial_phase_space_distribution}.
We first concentrate on the momentum part of this expression. Using integration by parts we transfer any derivative \wrt the $\tpi$ contained in the polynomial $\mathcal{C}$ from 
the Gaussian exponential over to the phase factor. All boundary terms vanish due to $\tens{C}_{pp}$ being positive definite as it defines a Gaussian distribution. We then have
\begin{eqnarray}
  & \usi{\tpi} \e^{\ii \, \tens{\bar{J}}_p \cdot \tpi } \mathcal{N}_p \, \mathcal{C}\left(\frac{\partial}{\ii\partial \tpi}\right) \e^{-\frac{1}{2}\tp^{(\ii)^{\T}} \tens{C}_{pp}^{-1} \, \tpi } \nonumber \\
 =& \usi{\tpi} \mathcal{C}\left(-\tens{\bar{J}}_p\right) \e^{\ii \, \tens{\bar{J}}_p \cdot \tpi } \mathcal{N}_p \, \e^{-\frac{1}{2} \tp^{(\ii)^{\T}} \tens{C}_{pp}^{-1} \, \tpi } 
 = \mathcal{C}\left(-\tens{\bar{J}}_p\right) \, \e^{-\frac{1}{2} \tens{\bar{J}}_p^{\T} \tens{C}_{pp} \, \tens{\bar{J}}_p} \;.
\label{eq:correlation_polynomial_partial_integration}
\end{eqnarray}
Using \eref{eq:time_averaged_Jp} and \eref{eq:momentum_gaussian_uncorrelated} we may rewrite the exponential in \eref{eq:correlation_polynomial_partial_integration} as
\begin{equation}
 \e^{-\frac{1}{2} \tens{\bar{J}}_p^{\T} \tens{C}_{pp} \, \tens{\bar{J}}_p} = \e^{-\frac{1}{2} \tens{\bar{J}}_p^{\T} \tens{\bar{C}}_{pp} \, \tens{\bar{J}}_p} \usi{\tpi} P_{\sigma_p^2}\left(\tpi\right) \e^{\ii \, \tens{\bar{J}}_p \cdot \tpi } \;.
\label{eq:restore_uncorrelated_momentum_integral}
\end{equation}
We can rewrite factors of $\tens{\bar{J}}_p$ in terms of functional derivatives by using
\begin{equation}
 \fd{}{\tK_p(t_\ii)} \e^{-\tJ \cdot \tens{\mathcal{G}} \cdot \tK} = \hat{\tchi}_p(t_\ii) \, \e^{-\tJ \cdot \tens{\mathcal{G}} \cdot \tK} = - \tens{\bar{J}}_p \, \e^{-\tJ \cdot \tens{\mathcal{G}} \cdot \tK} \;.
\label{eq:time_averaged_Jp_as_operator}
\end{equation}
With the definition of the correlation operator \eref{eq:total_corr_operator} we finally rewrite the free generating functional as
\begin{eqnarray}
  Z_{\CC,0}[\tJ,\tK] &= \usi{\txi} \CorOp\left(\hat{\tchi}_p(t_\ii)\right) \, \frac{P_{\sigma_p^2}\left(\tpi\right)}{V^N} \e^{\ii \left( \tens{\bar{J}}_q \cdot \tqi + \tens{\bar{J}}_p \cdot \tpi - \tJ \cdot \tens{\mathcal{G}} \cdot \tK \right)} \;,
\label{eq:free_gen_func_corr_operator_app}
\end{eqnarray}
which is equivalent to \eref{eq:free_gen_func_corr_operator} after rewriting the free solution in terms of path integrals.

\section{Correlation diagrams and cluster expansion} \label{app:diagrams_cluster}

We need to express the elementary building blocks of the correlation operator $\CorOp$ \eref{eq:total_corr_operator} in terms of line diagrams.
It will be easier to define at first only three line types, which while different from those in \eref{eq:corr_line_types} will have some notational overlap.
Once $\CorOp$ is expressed in terms of these three elementary line types we redefine the meaning of diagram lines as a combination of the elementary lines types
to arrive at \eref{eq:corr_line_types}. We begin by introducing
\begin{equation}
 \hat{C}_{p_j p_k} \coloneqq \nexp{ - \vochi_{p_j}^{\,\T}(t_\ii) \, C_{p_j p_k} \, \vochi_{p_k}(t_\ii) } - 1 = \mytikz{\hortwopattern[j][k] \ppline{p1}{p2} }
\label{eq:cpp_line}
\end{equation}
where Einstein summation is not implied. Note that the symmetry of the diagram reflects the symmetry of $C_{p_j p_k}$. 
The above term corresponds to a Mayer function $f_{ij} = \e^{-\beta v_{ij}}-1$ in the Mayer cluster expansion. We can now express the Gaussian in $\CorOp$ as
\begin{eqnarray}
\fl &\nexp{-\frac{1}{2} \, \hat{\tchi}_p^{\T}(t_\ii) \, \tens{\bar{C}}_{pp} \, \hat{\tchi}_p(t_\ii)} 
 = \nexp{- \sum_{\{i<j\}} \, \vochi_{p_i}^{\T}(t_\ii) \, C_{p_i p_j} \, \vochi_{p_k}(t_\ii)}  
 = \prod_{\{i<j\}} \left( 1 + \mytikz{\hortwopattern[i][j] \ppline{p1}{p2} } \right) \nonumber \\
\fl =& 1 + \sum_{\{i_1<j_1\}} \mytikz{\hortwopattern[i_1][j_1] \ppline{p1}{p2} } + \sum_{\{\{i_1<j_1\},\{i_2<j_2\}\}_{i_1<i_2}}  \mytikz{\hortwopattern[i_1][j_1] \ppline{p1}{p2} } \times \mytikz{\hortwopattern[i_2][j_2] \ppline{p1}{p2} } + \ldots 
\label{eq:momentum_gaussian_diagram_sum}
\end{eqnarray}
In the first line we used the symmetry of $C_{p_i p_j}$ to reduce the sum to ordered pairs of particles. When expanding out the product over them into sums we have to exclude identical pairs and
prevent overcounting, hence the ordering $i_1 < i_2$ between pairs. We now want to rewrite this not as a sum over multiple ordered pairs but over ordered $n$-tupels.
In the diagrammatic language this means we connect lines sharing a common particle label, \ie those with $j_n = j_m$ in the above case of \eref{eq:momentum_gaussian_diagram_sum}.
The fact that we only sum over ordered tupels of ordered pairs $i_n<i_m, \, \forall n<m$ translates into the 
\begin{description}
 \item[$p$-Line Rule:] Any pair of particles may at most be connected by one $\mytikz{\hortwopattern \ppline{p1}{p2} }$ line directly.
                This means that the subdiagram $\mytikz{ \hortwopattern \ppline{p1}{p2}[0.3cm] \ppline{p1}{p2}[-0.3cm]}$ is \emph{forbidden}.
\end{description}
Writing down any $n$-tupel sum in \eref{eq:momentum_gaussian_diagram_sum} then amounts to drawing and summing up all diagrams possible under this rule for a fixed graphical arrangement
of a general $n$-tupel of particle labels representing their ordering by ascending numerical value and then summing over all possible realisations of that tupel. This leads to
\begin{eqnarray}
  &\phantom{{}={}} \prod_{\{i<j\}} \left( 1 + \mytikz{\hortwopattern[i][j] \ppline{p1}{p2} } \right) =  1 + \sum_{n=2}^N \sum_{\{j_1 < \ldots < j_n\}} \Diag{n}|_{pp}(j_1,\ldots, j_n) \nonumber \\
  &= 1 + \sum_{\{i<j\}} \mytikz{\hortwopattern[i][j] \ppline{p1}{p2} } + \sum_{\{i<j<k\}} \left( \mytikz[2.0ex]{\threepattern[i][j][k] \ppline{p1}{p3} \ppline{p2}{p3} } + \mytikz[2.0ex]{\threepattern[i][j][k] \ppline{p1}{p2} \ppline{p1}{p3} } + \mytikz[2.0ex]{\threepattern[i][j][k] \ppline{p2}{p1} \ppline{p2}{p3} } + \mytikz[2.0ex]{\threepattern[i][j][k] \ppline{p1}{p2} \ppline{p2}{p3} \ppline{p3}{p1} }   \right) + \ldots
\label{eq:pp_line_sum}
\end{eqnarray}
where $N$ is the number of particles in the system. Notice that since we removed the momentum self-correlation from $\CorOp$ in \eref{eq:Cpp_split} no subdiagrams of the form
\mytikz[-0.1ex]{
 \coordinate[pdot] (p) at (0,0);
 \coordinate (center) at (0,0.2\gs);
 \draw[pline] circle[radius=0.2\gs, at=(center)];
}
appear in the above expression.

Turning to density auto-correlations and density-momentum cross-correlations we introduce diagrams
\begin{eqnarray}
 C_{\delta_j \delta_k} &\coloneqq \mytikz{\hortwopattern[j][k] \ddline{p1}{p2} } \;, \label{eq:Cdd_line} \\
 \hat{C}_{\delta_j p_k} &\coloneqq -\ii \vec{C}_{\delta_j p_k} \cdot \vochi_{p_k}(t_\ii) = \mytikz{\hortwopattern[j][k] \dpline{p1}{p2} }  \;. \label{eq:Cdp_line}
\end{eqnarray}
We can write the operator $\CorPol$ in $\CorOp$ following the same logic used in the step from \eref{eq:momentum_gaussian_diagram_sum} to \eref{eq:pp_line_sum},
\ie we multiply out all products over particle labels into sums of ordered tupels of labels by connecting lines at identical particle labels.
However, we need to take into account that the form \eref{eq:initial_correlation_polynomial} of $\CorPol$ imposes certain restrictions on the possible diagrams.
\begin{itemize}
 \item In any term of the ordered sums $\sum_{\{\{i_1<j_1\},\ldots,\{i_n<j_n\}\}'_{i_1<\ldots<i_n}}$ over the labels of the $C_{\delta_{i_k} \delta_{j_k}}$ any particle
       label may only appear once. We may thus not connect two $\mytikz{ \hortwopattern \ddline{p1}{p2} }$-lines with one another.
 \item In the products $\prod_{\{n\}'} \left( 1 + \sum_{m \neq n} \hat{C}_{\delta_n p_m} \right)$ each label $n$ pertaining to $\delta_n$ only appears once. We may thus not connect
       two $\mytikz{ \hortwopattern \dpline{p1}{p2} }$-lines with their solid $\delta$ end.
 \item The above products do not share labels with the above sums, as indicated by the prime. This forbids connecting the solid $\delta$ end of a $\mytikz{ \hortwopattern \dpline{p1}{p2} }$-line to a
       $\mytikz{ \hortwopattern \ddline{p1}{p2} }$-line.
\end{itemize}
These three restrictions can be combined into the $\delta$-line rule given in section \ref{sec:cluster}, albeit that for now it holds for the three elementary line types.
Together with the $m \neq n$ restriction in the second of the above points the $\delta$-line rule also excludes any form of self-correlations, \ie one-particle loops.
The $\CorPol$-operator may thus be written as
\begin{eqnarray}
 \CorPol\left(\hat{\tchi}_p(t_\ii)\right) &= 1 + \sum_{n=2}^N \sum_{\{j_1 < \ldots < j_n\}} \Diag{n}|_{\delta\delta,\delta p}(j_1,\ldots, j_n) \nonumber \\
 &= 1 + \sum_{\{i<j\}} \left( \mytikz{\hortwopattern[i][j] \ddline{p1}{p2} } +
                              \mytikz{\hortwopattern[i][j] \dpline{p1}{p2} } +
                              \mytikz{\hortwopattern[i][j] \dpline{p2}{p1} } +
                              \mytikz{\hortwopattern[i][j] \ppline{p1}{p2} } +
                              \mytikz{\hortwopattern[i][j] \dpline{p1}{p2}[0.3cm] \dpline{p2}{p1}[0.3cm] } \right) \nonumber \\
 &\phantom{{}={}} + \sum_{\{i<j<k\}} \left( \mytikz[2.0ex]{\threepattern[i][j][k] \dpline{p1}{p3} \dpline{p2}{p3}} +
                                            \mytikz[2.0ex]{\threepattern[i][j][k] \dpline{p2}{p1} \dpline{p3}{p1}} +
                                            \mytikz[2.0ex]{\threepattern[i][j][k] \dpline{p1}{p2} \dpline{p3}{p2}} +
                                            \ldots \right) + \ldots
\label{eq:dd_dp_line_sum}
\end{eqnarray}

The last step is to multiply $\CorPol$ \eref{eq:dd_dp_line_sum} with the Gaussian exponential \eref{eq:pp_line_sum}. Since there are no restrictions on connecting
$\mytikz{\hortwopattern \ppline{p1}{p2} }$ with either $\mytikz{\hortwopattern \ddline{p1}{p2} }$ or $\mytikz{\hortwopattern \dpline{p1}{p2} }$, we simply obtain $\CorOp$
as the sum over all diagrams that can be constructed from the three line types subject to both the $\delta$-line and $p$-line
rules. We illustrate this by writing it out for the two-particle contribution
\begin{eqnarray}
\CorOp &= 1 + \sum_{n=2}^N \sum_{\{j_1 < \ldots < j_n\}} \Diag{n}(j_1,\ldots, j_n) \nonumber \\
                                 &= 1 + \sum_{\{i<j\}} \left( \mytikz{\hortwopattern[i][j] \ddline{p1}{p2} } + \mytikz{\hortwopattern[i][j] \dpline{p1}{p2} } + \mytikz{\hortwopattern[i][j] \dpline{p2}{p1} } + \mytikz{\hortwopattern[i][j] \ppline{p1}{p2} } + \mytikz{\hortwopattern[i][j] \dpline{p1}{p2}[0.3cm] \dpline{p2}{p1}[0.3cm] } \right. \nonumber \\
                                 &\phantom{{}={}} \left. + \mytikz{\hortwopattern[i][j] \ddline{p1}{p2}[0.3cm] \ppline{p1}{p2}[-0.3cm] } + \mytikz{\hortwopattern[i][j] \dpline{p1}{p2}[0.3cm] \ppline{p1}{p2}[-0.3cm] } + \mytikz{\hortwopattern[i][j] \dpline{p2}{p1}[-0.3cm] \ppline{p1}{p2}[-0.3cm] } + \mytikz{\hortwopattern[i][j] \dpline{p1}{p2}[0.3cm] \dpline{p2}{p1}[0.3cm] \ppline{p1}{p2} } \right) + \sum_{\{i<j<k\}} \ldots
\label{eq:Ctot_two_particles}
\end{eqnarray}
In this form of the diagrammatic language the complete three-particle contribution would already contain a plethora of diagrams with up to six lines.
We can remedy this to some degree by recognizing the trivial fact that the diagrams in \eref{eq:Ctot_two_particles} are also the building blocks of all diagrams with three particles or more.
By combining diagrams and thus reducing their number in \eref{eq:Ctot_two_particles} we effectively do so in all higher order terms as well. The form of \eref{eq:Ctot_two_particles} suggests
the \textit{diagram redefinition}
\begin{eqnarray}
 \mytikz{\hortwopattern[i][j] \ddline{p1}{p2} } + \mytikz{\hortwopattern[i][j] \ddline{p1}{p2}[0.3cm] \ppline{p1}{p2}[-0.3cm] } \; \rightarrow \; \mytikz{\hortwopattern[i][j] \ddline{p1}{p2} } &= C_{\delta_i \delta_j} \, \e^{-\vochi_i^{\,\T}(t_\ii) \, C_{p_i p_j} \, \vochi_j(t_\ii)} \;, \nonumber \\
 \mytikz{\hortwopattern[i][j] \dpline{p1}{p2} } + \mytikz{\hortwopattern[i][j] \dpline{p1}{p2}[0.3cm] \ppline{p1}{p2}[-0.3cm] } \; \rightarrow \; \mytikz{\hortwopattern[i][j] \dpline{p1}{p2} } &= \hat{C}_{\delta_i p_j} \, \e^{-\vochi_i^{\,\T}(t_\ii) \, C_{p_i p_j} \, \vochi_j(t_\ii)} \;, \nonumber \\
 \mytikz{\hortwopattern[i][j] \dpline{p1}{p2}[0.3cm] \dpline{p2}{p1}[0.3cm] } + \mytikz{\hortwopattern[i][j] \dpline{p1}{p2}[0.3cm] \dpline{p2}{p1}[0.3cm] \ppline{p1}{p2} } \; \rightarrow \; \mytikz{\hortwopattern[i][j] \ddlline{p1}{p2} } &= \hat{C}_{\delta_i p_j} \, \hat{C}_{\delta_j p_i} \, \e^{-\vochi_i^{\,\T}(t_\ii) \, C_{p_i p_j} \, \vochi_j(t_\ii)} \;.
 \label{eq:diagram_redefinition}
\end{eqnarray}
The meaning of the $\mytikz{\hortwopattern \ppline{p1}{p2} }$ line is unchanged.
It is straightforward to see by checking all possible combinations that for these new diagrams the former $\delta$-line rule and $p$-line rule combine into the same $\delta$-line rule
and a general rule forbidding two-particle loops. Together with the exclusion of one-particle loops these are the three diagram rules given in section \ref{sec:cluster}.
The operator $\CorOp$ is then expressed as in \eref{eq:Ctot_diag_sum}, where the diagram sum now includes all diagrams that can be drawn with the redefined line types
under the three restriction rules.

In order to arrive at \eref{eq:free_gen_func_factorised} we now need to reorder $\CorOp$ and through it the free generating functional $Z_{\CC,0}$ \eref{eq:free_gen_func_corr_operator}
in terms of connected diagrams, \ie clusters. We already introduced the notion of a cluster configuration $\{m_\ell\}$ in section \ref{sec:cluster} as the collection of all diagrams
made up of $m_\ell$ clusters of $\ell$ particles for all $\ell = 1,\ldots,N$. Any individual diagram corresponds to an \textit{assignment} of the $N$ particles
to the different clusters. For the $\{m_3=1, m_2=1, m_1=N-5\}$ configuration, a possible assigment would be $[(1,2,3),(4,5),(6),\ldots,(N)]$.
Any such assigment is realised by many different products of connected diagrams, in the above case one of them would be
\begin{eqnarray}
  &\left( \prod_{j=1}^N \Tr_j \right) \; \mytikz[2.0ex]{\threepattern[1][2][3] \ppline{p1}{p2} \ppline{p1}{p3} } \; \mytikz[2.0ex]{\vertwopattern[4][5] \ddline{p1}{p2} } \; \prod_{j=6}^N \mytikz{\pdot[j]} \nonumber \\
 =& \Tr_1 \Tr_2 \Tr_3  \mytikz[2.0ex]{\threepattern[1][2][3] \ppline{p1}{p2} \ppline{p1}{p3} }  \; \Tr_4 \Tr_5 \mytikz[2.0ex]{\vertwopattern[4][5] \ddline{p1}{p2} } \; \prod_{j=6}^N \Tr_j \mytikz{\pdot[j]} \;.
 \label{eq:cluster_config_example}
\end{eqnarray}
where we represented the non-correlated particles by unconnected dots, \ie 1-clusters representing factors of unity.
Another possibility to realise this assignment can be found by replacing the dashed line between particles $1$
and $2$ with a solid line. Adding the result to \eref{eq:cluster_config_example} we obtain
\begin{equation}
\Tr_1 \Tr_2 \Tr_3 \left( \mytikz[2.0ex]{\threepattern[1][2][3] \ppline{p1}{p2} \ppline{p1}{p3} } + \mytikz[2.0ex]{\threepattern[1][2][3] \ddline{p1}{p2} \ppline{p1}{p3} } \right) \; \Tr_4 \Tr_5 \mytikz[2.0ex]{\vertwopattern[4][5] \ddline{p1}{p2} } \; \prod_{j=6}^N \Tr_j \mytikz{\pdot[j]} \;. 
\label{eq:cluster_config_example_diagram_added}
\end{equation}
We now add all other possible connected diagrams for the three-particle cluster $(1,2,3)$ to \eref{eq:cluster_config_example_diagram_added} while keeping the rest of the diagram fixed. Afterwards
we do the same for the two-particle cluster $(4,5)$ and, using the definition of the connected $\ell$-particle generating functional \eref{eq:l_cluster_gen_func}, arrive at
\begin{equation}
 \left( 3! \, \cgf{3} \right) \, \left( 2! \, \cgf{2} \right) \, \left( 1! \, \cgf{1} \right)^{N-5} \;.
\label{eq:cluster_config_example_factorised}
\end{equation}
From this we directly see that the contribution of any particle assignment in a cluster configuration $\{m_\ell\}$ to the free generating functional \eref{eq:free_gen_func_trace}
may be written as
\begin{equation}
 \prod_{\ell=1}^N \left(\ell! \, \cgf{\ell}\right)^{m_\ell} \;.
\label{eq:cluster_config_example_general_factorisation}
\end{equation}

The last step is to figure out how many particle assignments are actually present in \eref{eq:free_gen_func_trace} per given cluster configuration. By simply rearranging the particle
order freely in \eg $[(1,2,3),(4,5),(6),\ldots,(N)]$, there are in principle $N!$ possible assignments for any $\{m_\ell\}$. However, we defined the correlation operator
\eref{eq:Ctot_two_particles} as a sum over ordered tupels of labels. Once we have fixed which particles are clustered together we must thus only include one assignment realising this clustering.
We therefore divide out all $\prod_{\ell=1}^N \ell!^{m_\ell}$ reorderings of particle labels inside the clusters and all $\prod_{\ell=1}^N m_\ell!$
ways to exchange all particles between two clusters of equal size. This finally leads to the expression \eref{eq:free_gen_func_factorised} for the free generating functional
\begin{eqnarray}
  Z_{\CC,0} &= \sum_{\{m_\ell\}^\ast} \frac{N!}{\prod_{\ell=1}^N (\ell!)^{m_\ell} m_\ell!} \prod_{\ell=1}^N (\ell!W_0^{(\ell)})^{m_\ell} = N! \sum_{\{m_\ell\}^\ast} \prod_{\ell=1}^N \frac{ \left(W_0^{(\ell)}\right)^{m_\ell}} {m_\ell!} \;,
\label{eq:free_gen_func_cluster_derivation}
\end{eqnarray}
where the asterisk in the sum represents the constraint \eref{eq:cluster_configuration_constraint}.

\section{Operator evaluation and homogeneity restriction} \label{app:op_eval_hom_restriction}

In order to calculate general cumulants \eref{eq:general_cumulant_operator_product} we need to evaluate the collective field operators $\PhiOp{\ell}_{\alpha_r}(r)$
and the correlation operators \eref{eq:corr_line_types} represented by the diagrams. We first introduce the unaveraged $\ell$-particle free generating functional
\begin{equation}
 \unavZ{\ell}\left[\tJ,\tK\right]
 \coloneqq \bpint{\tx}{\txi}{} \upi{\tchi} \e^{\ii \left( \tchi \cdot \left(\partial_t + \tens{F}\right) \tx + \tJ \cdot \tx + \tK \cdot \tchi \right)}
 = \e^{\ii \, \tJ \cdot \bar{\tx}} \;.
\label{eq:unaveraged_gen_func}
\end{equation}
We write out the product of collective field operators in \eref{eq:general_cumulant_operator_product} into individual terms and in each of them pull all
one-particle phase-space density operators to the right. As already shown in \cite{Bartelmann2016}, the effect of applying such an operator is
\begin{eqnarray}
 \PhiOp{1}_{f_j}(1) \, \unavZ{\ell}[\tJ,\tK] = \nexp{\ii \bsi{t}{t_\ii}{t_\ff} \left(- \dirac\left(t-t_1\right) \vs_1 \otimes \ve_j \right)^{\,\T} \bar{\tx}(t) } \e^{ \ii \, \tJ \cdot \bar{\tx} } \;.
\label{eq:one_particle_density_operator_evaluation}
\end{eqnarray}
This is easily extended to the case where the particle $j$ carries some subset $\II_j$ of the field labels $1,\ldots,n$ by defining the following phase shift vector
\begin{equation}
\tL_j(t) \coloneqq - \sum_{r \in \II_j} \dirac\left(t - t_r\right) \vec{s}_r \otimes \vec{e}_j \;,
\label{eq:pre_phase_shift_vector}
\end{equation}
which allows us to write the effect of applying all single-particle phase-space density operators in some term of \eref{eq:general_cumulant_operator_product} as a shift of the $\tJ$ source
\begin{equation}
\unavZ{\ell}[\tJ+\tL,\tK] \coloneqq \unavZ{\ell}[\tJ + \sum_{j=1}^\ell \tL_j(t),\tK] = \left( \prod_{j=1}^\ell \left( \prod_{r \in \II_j} \PhiOp{1}_{f_j}(r) \right) \right) \unavZ{\ell}[\tJ,\tK]  \;.
\label{eq:04-02-08}
\end{equation}
This shows that knowledge of $\tL$ is equivalent to the explicit grouping of field labels into sets $\II_j$ carried by particles $j$.
Assume for now that all functional derivatives \wrt $\tK_p(t_\ii)$ have been executed, which leaves no operators remaining and we may therefore set all sources $\tJ$ and $\tK$ to zero.
For the unaveraged generating functional we then find
\begin{eqnarray}
\unavZ{\ell}[\tJ+\tL,\tK] \Big|_0  &= \unavZ{\ell}[\tL,\tens{0}] = \nexp{-\ii \bsi{t}{t_\ii}{t_\ff} \sum_{j=1}^\ell \sum_{r \in \II_j} \dirac\left(t - t_r\right) \vs_r^{\,\T} \mathcal{G}(t,t_\ii) \, \vx_j^{\,(\ii)} } \nonumber \\ 
                                                  &= \prod_{j=1}^{\ell} \, \e^{-\ii \left( \Lq{j}(t_\ii) \cdot \vqi_j + \Lp{j}(t_\ii) \cdot \vpi_j \right) }  \;,
\label{eq:unaveraged_gen_func_shifted}
\end{eqnarray}
where the one-particle \textit{phase shift vectors} $\Lq{j}, \Lp{j}$ have been defined in \eref{eq:q_shift_vector}, \eref{eq:p_shift_vector}.

Both the single particle response operators $\hat{b}_j$ and all correlation operators are defined purely in terms of operators $\vochi_{p_j}$. For a given shift vector $\tL$ they evaluate to
\begin{eqnarray}
   & \vochi_{p_j}(t_u) \, \unavZ{\ell}[\tJ+\tL,\tK] \Big|_0 = \fd{}{\vec{K}_{p_j}(t_u)} \, \e^{\ii \left(\tJ + \tL\right) \cdot \left( \tens{\mathcal{G}}\tx^{(\ii)} -\tens{\mathcal{G}} \cdot \tK\right)} \Big|_0 \nonumber \\
  =& \left[ -\bsi{t}{t_\ii}{t_\ff} \Big(- \sum_{k=1}^\ell \sum_{r \, \in \, \II_k} \, \dirac\left(t - t_r\right) \vs_r \otimes \ve_k \Big)^{\T} \left( \mathcal{G}(t,t_u) \cvector{0 \\ \mathcal{I}_d } \otimes \ve_j \right) \right]^{\T} \unavZ{\ell}[\tL,0] \nonumber \\
  =& \Lp{j}(t_u) \, \unavZ{\ell}[\tL,\tens{0}] \;.
\label{eq:mom_shift_operator_evaluation}
\end{eqnarray}
The physical interpretation of $\vochi_{p_j}$ is thus that it generates the phase shift vector which when multiplied with some momentum perturbation $\Delta\vec{p}$ 
gives the linear response of the system in form of a Fourier phase shift. As argued in section \ref{sec:ini_corr_operator}, the initial correlations introduce such momentum perturbations relative to
the uncorrelated system. The correlation operators represented by lines \eref{eq:corr_line_types} in correlation diagrams thus evaluate to
\begin{eqnarray}
 \mytikz{\hortwopattern[j][k] \ppline{p1}{p2}} \unavZ{\ell}[\tJ+\tL,\tK] \Big|_0 &= \hat{C}_{p_i p_j}\left(\vqi_{ij}, \Lp{j},\Lp{k}\right) \unavZ{\ell}[\tL,\tens{0}] \;. 
\label{eq:corr_line_evaluated}
\end{eqnarray}
and analogously for the three other line types. We now understand $\CDiag{\ell}(\tL)$ as the sum over all $\ell$-particle diagrams with lines evaluated in the sense of
\eref{eq:corr_line_evaluated}. A general $\ell$-particle phase-space density cumulant is then written as
\begin{equation}
 \Gf[\ell]_{f(1) \cdots f(n)} = \mpd^\ell \sum_{\tL} \usi{\tx^{(\ii)}} P_{\sigma_p^2}\left(\tens{p}^{(\ii)}\right) \unavZ{\ell}[\tL,\tens{0}] \, \frac{1}{\ell!} \, \CDiag{\ell}(\tL) \;.
\label{eq:general_f_cumulant_L_sum}
\end{equation}

Note that so far the sum over all possible $\tL$ also includes all cases where label sets may be empty.
We now show how the homogeneity restriction of section \ref{sec:general_form_cumulants} comes about by 
picking any term, where without loss of generality particle $j$ carries no field label, \ie $\II_j = \emptyset$. Since particle $j$ is involved in the cumulant
it must appear in the diagrams of $\CDiag{\ell}(\tL)$ and be connected with one of the four line types shown in \eref{eq:corr_line_types}. 
If the particle is connected by more than one line, then the $\delta$-line rule demands that at most one of the lines connect to $j$ with a
dashed $p$-side. But since $\II_j = \emptyset$ we have $\Lp{j} = 0$ and by combining \eref{eq:corr_line_evaluated} with \eref{eq:corr_line_types}
we see that in this case any corresponding correlation factor is zero. If the particle is connected to only one line, we need to distinguish three different cases
to complete the proof:
\begin{itemize}
  \item The particle is connected to a dashed $p$-type line so the same argument as above applies.  
  \item It is connected to a $C_{\delta_j \delta_k}$ line. Since $\II_j = \emptyset$ we also have $\Lq{j} = 0$ and according to \eref{eq:unaveraged_gen_func_shifted}
        \begin{equation*}
         \unavZ{\ell}[\tL,\tens{0}] \propto \e^{-\ii \, \Lq{j}(t_\ii) \cdot \vqi_j } = \e^0 = 1 \;.
        \end{equation*}
        Thus the only remaining quantity in the term that depends on $\vqi_j$ is $C_{\delta_j \delta_k}$. Using the definition of the density contrast gives
        \begin{equation*}
         \int_{\qin_j} C_{\delta_j \delta_k} = \ave{ \int_{\qin_j} \delta^{(\ii)}(\vqi_j) \, \delta^{(\ii)}(\vqi_k)} = 0 \;.
        \end{equation*}             
  \item It is connected to the solid $\delta$-side of a $\hat{C}_{\delta_j p_k}$. The same arguments as in the previous case apply and we find
        \begin{equation*}
          \int_{\qin_j} \vec{C}_{\delta_j p_k} = \ave{  \int_{\qin_j} \delta^{(\ii)}(\vqi_j) \, \vec{P}^{\,(\ii)}(\vqi_k)} = 0 \;.
        \end{equation*}
\end{itemize}

The homogeneity restriction excludes all shift vectors $\tL$ corresponding to terms with empty label sets from the sum in \eref{eq:general_f_cumulant_L_sum}.
We now make the change of perspective to label groupings where we only specify which labels are grouped into $\ell$ non-empty sets enumerated as $\II_m$ without assigning them to particles $j$.
We use the convention that writing the sets in the order $(\II_1, \ldots, \II_\ell)$ is equivalent to using the identity mapping between set labels $m$ and particle labels $j$,
\ie particle $j$ carries set $\II_m$ with $m=j$. All other possible $\ell!$ assignments of this specific grouping are then obtained by all permutations of the order of the sets.
At this point the particle labels are essentially redundant information, so we may always understand labels $j$ as pertaining to the sets of $\II_j$ of the grouping.

The sum over $\tL$ in \eref{eq:general_f_cumulant_L_sum} can then be written as summing over all possible label groupings
and then summing each of them over all permutations of ordering the label sets.
With the exception of $\CDiag{\ell}$ all other quantities in \eref{eq:general_f_cumulant_L_sum} are invariant under set order permutations.
We now use the fact that $\CDiag{\ell}$ does not depend on the initial momenta of the particles to execute the integration over them. By combining the momentum part of
\eref{eq:unaveraged_gen_func_shifted} with \eref{eq:momentum_gaussian_uncorrelated}, this leads to the Gaussian damping factor
\begin{eqnarray}
\e^{-Q(\II_1,\ldots,\II_\ell)} &\coloneqq \usi{\tpi} P_{\sigma_p^2}\left( \tpi \right) \prod_{j=1}^\ell \e^{ -\ii  \Lp{j}(t_\ii) \cdot \vpi_j } \nonumber \\
                               &= \prod_{j=1}^\ell \int_{\pin_j} P_{\sigma_p^2}\left(\vpi_j\right) \, \e^{ -\ii  \Lp{j}(t_\ii) \cdot \vpi_j } 
                                = \e^{ -\frac{\sigma_p^2}{2} \sum_{j=1}^\ell \left(\Lp{j}(t_\ii)\right)^2 } \;,
\label{eq:damping_factor_derivation}
\end{eqnarray}
where we factorised the uncorrelated Gaussian distribution into its individual particle contributions in the second step. At this point we may write the general phase-space density cumulant as
\begin{eqnarray}
 \Gf[\ell]_{f(1) \cdots f(n)} &= \mpd^\ell \sum_{\{\II_1,\ldots,\II_\ell\}} \e^{-Q(\II_1,\ldots,\II_\ell)} \left(\prod_{j=1}^\ell \int_{\qin_j} \e^{ -\ii  \Lq{j}(t_\ii) \cdot \vqi_j } \right) \nonumber \\
                                 &\phantom{{}={}} \frac{1}{\ell!} \, \sum_{\pi} \CDiag{\ell}\left(\II_{\pi(1)},\ldots,\II_{\pi(\ell)}\right) \;.
\label{eq:general_f_cumulant_set_sum}
\end{eqnarray}

\section{Derivation of Feynman rules} \label{app:feynman_rules_derivation}

We first need to show that the sum over permutations of field-label sets in \eref{eq:general_f_cumulant_set_sum} cancels against the preceding factor of $1/\ell!$.
For this it is advantageous to go back to the definition \eref{eq:l_cluster_gen_func} of $\cgf{\ell}$ and classify the connected diagrams in terms of an equivalence relation.
For a given representative diagram of some fixed ordered $n$-tupel of \textit{particle} labels $\{j_1 < \ldots < j_n\}$ we define the sum over its \textit{permutation equivalence class}
as the sum of all diagrams that can be transformed into the representative under some permutation of these particle labels. For the first three diagrams in \eref{eq:Ctot_pp_three_particles} this would be
\begin{eqnarray}
 \left[ \mytikz[2.0ex]{\threepattern[i][j][k] \ppline{p1}{p3} \ppline{p2}{p3} } \right] 
 &\coloneqq \mytikz[2.0ex]{\threepattern[i][j][k] \ppline{p1}{p3} \ppline{p2}{p3} } + \mytikz[2.0ex]{\threepattern[i][j][k] \ppline{p1}{p2} \ppline{p1}{p3} } + \mytikz[2.0ex]{\threepattern[i][j][k] \ppline{p2}{p1} \ppline{p2}{p3} } \nonumber \\
 &= \mytikz[2.0ex]{\threepattern[i][j][k] \ppline{p1}{p3} \ppline{p2}{p3} } + \mytikz[2.0ex]{\threepattern[j][k][i] \ppline{p1}{p3} \ppline{p2}{p3} } + \mytikz[2.0ex]{\threepattern[k][i][j] \ppline{p1}{p3} \ppline{p2}{p3} } \eqqcolon \sum_{\mathrm{Perm}(i,j,k)\diagup \mathrm{Sym.}} \mytikz[2.0ex]{\threepattern[i][j][k] \ppline{p1}{p3} \ppline{p2}{p3} }
\end{eqnarray}
The second line shows that we could also define the sum over the equivalence class as the sum of the representative diagram over all possible permutations
of particle labels modulo those permutations which leave the diagram invariant, \ie those which define the symmetry of the diagram.
This particular diagram has a symmetry factor of $\mathcal{S}=2$ which leads us to exclude the diagrams
\begin{equation}
 \mytikz[2.0ex]{\threepattern[j][i][k] \ppline{p1}{p3} \ppline{p2}{p3} } = \mytikz[2.0ex]{\threepattern[i][j][k] \ppline{p1}{p3} \ppline{p2}{p3} } \;, \quad
 \mytikz[2.0ex]{\threepattern[k][j][i] \ppline{p1}{p3} \ppline{p2}{p3} } = \mytikz[2.0ex]{\threepattern[j][k][i] \ppline{p1}{p3} \ppline{p2}{p3} } \;, \quad
 \mytikz[2.0ex]{\threepattern[i][k][j] \ppline{p1}{p3} \ppline{p2}{p3} } = \mytikz[2.0ex]{\threepattern[k][i][j] \ppline{p1}{p3} \ppline{p2}{p3} }
 \label{eq:symmetric_diagram_exclusion}
\end{equation}
in order to avoid overcounting. In the original definition of the permutation equivalence class this overcounting was prevented by the fixed order of the particle labels.
In general there are $\ell!$ possible permutations of a given ordered $\ell$-tuple of particle labels. Consider some $\ell$-particle diagram representative $\mathcal{R}$, then it must hold
that $\ell! = \mathcal{S}(\mathcal{R}) \, N_{[\mathcal{R}]}$ for the number $N_{[\mathcal{R}]}$ of diagrams in its equivalence class.

Since the $\ell$-particle trace is invariant under particle label permutations, all diagrams in a permutation equivalence class give the same contribution to $\cgf{\ell}$.
For each permutation equivalence class we may choose a representative, giving us a set $\mathcal{R}^{(\ell)}_a$, $a = 1,\ldots,M^{(\ell)}_{\sim}$.
In terms of these we can write the $\ell$-cluster generating functional as
\begin{equation}
 \cgf{\ell} = \frac{1}{\ell!} \, \left( \prod_{j=1}^\ell \Tr_j \right) \sum_{a=1}^{M^{(\ell)}_{\sim}} \left[\mathcal{R}^{(\ell)}_a(1,\ldots,\ell)\right] = \left( \prod_{j=1}^\ell \Tr_j \right) \sum_{a=1}^{M^{(\ell)}_{\sim}} \frac{\mathcal{R}^{(\ell)}_a(1,\ldots,\ell)}{\mathcal{S}(\mathcal{R}^{(\ell)}_a)} \;.
\label{eq:cluster_gen_func_repr_sum}
\end{equation}
Since our expression \eref{eq:general_f_cumulant_set_sum} for the general phase-space density cumulant is defined in terms of $\cgf{\ell}$ we may insert \eref{eq:cluster_gen_func_repr_sum} to find
\begin{eqnarray}
    \frac{1}{\ell!} \, \sum_{\pi} \CDiag{\ell}\left(\II_{\pi(1)},\ldots,\II_{\pi(\ell)}\right)
    &= \sum_{a=1}^{M^{(\ell)}_{\sim}} \sum_{\pi} \frac{\mathcal{R}^{(\ell)}_a\left(\II_{\pi(1)},\ldots,\II_{\pi(\ell)}\right)}{\mathcal{S}(\mathcal{R}^{(\ell)}_a)} \nonumber \\
    &= \CDiag{\ell}\left(\II_1,\ldots,\II_\ell\right) \;.
\label{eq:permutation_sum}
\end{eqnarray}
The second step uses the fact that summing the representative over all permutations results in the sum over the entire equivalence class,
but without excluding symmetric terms like in \eref{eq:symmetric_diagram_exclusion}. These cancel against the symmetry factors.

The fact that due to statistical homogeneity initial correlations between particles only depend on their relative distance has been exploited in
\cite{Bartelmann2017} to factorize correlator contributions into products of a general quantity $\mathcal{P}$, which is essentially given
by the Fourier transform of the $\mytikz{\hortwopattern \ppline{p1}{p2}}$-line. We repeat the same logic in the present case where we consider all types of correlations and not only
momentum correlations. Without loss of generality we shift the origin of our coordinate system to the initial position $\vqi_\ell$ leading to new coordinates 
\begin{equation}
 \vqi_{j\ell} \coloneqq \vqi_{j} - \vqi_{\ell} \qquad \forall \; j \in \{1,\ldots,\ell-1\} \;.
\label{eq:centered_position_vectors}
\end{equation}
Relative distance vectors are then given by
\begin{equation}
 \vqi_{jk} = \vqi_{j\ell} - \vqi_{k\ell} \quad \mathrm{where} \quad j,k < \ell \;.
\label{eq:relative_position_vectors}
\end{equation}
Since $\CDiag{\ell}$ only depends on the $\vqi_{jk}$ we can execute the integration over $\vqi_\ell$ to find
\begin{eqnarray}
   \prod_{j=1}^{\ell} \int_{\qin_j} \, \e^{-\ii \Lq{j}(t_\ii) \cdot \vqi_j } 
 = (2\pi)^d \dirac\left( \sum_{j=1}^\ell \Lq{j}(t_\ii) \right) \prod_{j=1}^{\ell-1} \int_{\qin_{j\ell}} \, \e^{-\ii \Lq{j}(t_\ii) \cdot \vqi_{j \ell}}  \;.
\label{eq:q_ell_integration}
\end{eqnarray}
Inserting \eref{eq:permutation_sum} and \eref{eq:q_ell_integration} into \eref{eq:general_f_cumulant_set_sum} then leads to the general form
\eref{eq:general_cumulant_sum_groupings} for a phase-space density cumulant. By introducing appropriate Dirac delta distributions that enforce the constraints \eref{eq:relative_position_vectors}
and rewriting them in terms of their Fourier transforms we can extend the integration to all relative position vectors
\begin{eqnarray}
   \prod_{j=1}^{\ell-1} \int_{\qin_{j\ell}} \, \e^{-\ii \Lq{j}(t_\ii) \cdot \vqi_{j \ell}} 
 = \left( \prod_{(a<b<\ell)} \int_{\kin_{ab}} \right) \prod_{j<k} \int_{\qin_{jk}} \, \e^{-\ii \Ki_{jk} \cdot \vqi_{jk}} \;.
\label{eq:q_jk_integration_extension}
\end{eqnarray}
This introduces Fourier vectors $\Ki_{jk}$ conjugate to the $\vqi_{jk}$ with $j<k$ which are defined as
\begin{equation}
\Ki_{jk} \coloneqq
 \cases{ \Lq{j}(t_\ii) + \sum_{a=1}^{j-1} \vki_{aj} - \sum_{b=j+1}^{\ell-1} \vki_{jb} & for $k=\ell, j \in \{1,\ldots,\ell-1\}$ \\
         \vki_{jk} & else. \\}
\label{eq:generalised_fourier_momenta}
\end{equation}
Applying these Fourier transforms to $\CDiag{\ell}$ then allows us to transform each correlation line independently which leads to the general definition of Fourier
transformed correlation lines given in \eref{eq:fourier_corr_lines}, where $\mathcal{P}_{p_j p_k}$ corresponds to the $\mathcal{P}_{jk}$ already discussed in \cite{Bartelmann2017}.

One can easily check that the Fourier momenta \eref{eq:generalised_fourier_momenta} obey the conservation law \eref{eq:fourier_momentum_conservation}. We first consider
the case $j \neq \ell$.
\begin{eqnarray}
  & \Lq{j} + \sum_{i=1}^{j-1} \Ki_{ij} - \sum_{i=j+1}^\ell \Ki_{ji} = \Lq{j} + \sum_{i=1}^{j-1} \Ki_{ij} - \sum_{i=j+1}^{\ell-1} \Ki_{ji} - \Ki_{j\ell} \nonumber \\
 =& \Lq{j} + \sum_{i=1}^{j-1} \vki_{ij} - \sum_{i=j+1}^{\ell-1} \vki_{ji} - \left( \Lq{j} + \sum_{a=1}^{j-1} \vki_{aj} - \sum_{b=j+1}^{\ell-1} \vki_{jb} \right) = 0 
\label{eq:momentum_conservation_proof_a}
\end{eqnarray}
The case $j = \ell$ is only slightly more complicated. Since $\Ki_{\ell i} = 0$ by definition we find
\begin{eqnarray}
  & \Lq{\ell} + \sum_{i=1}^{\ell-1} \Ki_{i\ell} = \Lq{\ell} + \sum_{i=1}^{\ell-1} \left( \Lq{i} + \sum_{a=1}^{i-1} \vki_{ai} - \sum_{b=i+1}^{\ell-1} \vki_{ib} \right) \nonumber \\
 =& \sum_{i=1}^{\ell} \Lq{i} + \sum_{1 \leq a < i \leq \ell-1} \vki_{ai} - \sum_{1 \leq i < b \leq \ell-1} \vki_{ib} = 0 
\label{eq:momentum_conservation_proof_b}
\end{eqnarray}
The Feynman rules for assigning momenta follow from the fact that any open pair $\mytikz{ \hortwopattern[j][k]}$ of particle vertices $j<k$ in a diagram of $\CDiag{\ell}$ leads to
a factor
\begin{equation}
 \int_{\qin_{jk}} \, \e^{-\ii \Ki_{jk} \cdot \vqi_{jk}} = (2\pi)^d \dirac\left(\Ki_{jk}\right) \;.
\label{eq:open_line_dirac}
\end{equation}
We can first go through all open pairs where $k \neq \ell$ and thus $\Ki_{jk} = \vki_{jk}$. For these we can execute the appropriate $\int_{\kin_{jk}}$
from \eref{eq:q_jk_integration_extension} to obtain a factor of unity and drop $\vki_{jk}$ from all conservation laws.
For any remaining open pairs $j<k$ with $k = \ell$ we can choose and execute one of the remaining $\int_{\kin_{ab}}$ to replace $\vki_{ab}$ in all conservation laws
with the constraint obtained from solving the equation $\Ki_{j\ell} = 0$ for $\vki_{ab}$.
Once all Dirac delta distributions of the form \eref{eq:open_line_dirac} have been exploited, some non-trivial $\vki_{ab}$-integrations will remain as the loop momenta
of our Feynman-rules, \ie their choice is equivalent to the choice of the $\vki_{ab}$-integrations we use to resolve the $\dirac(\Ki_{j\ell})$.

To see that this conforms to the familiar idea of loops in Feynman diagrams, consider the number of particle pairs $N_{\mathrm{pairs}}$,
the number of lines $P$ playing the role of Feynman-propagators, the number of particle vertices $V=\ell$ and the total number $I$ of $\vki_{ab}$-integrals we had to introduce.
If $L$ is the number of non-trivial $\vki_{ab}$-integrals we find
\begin{equation}
 L = I - (N_{\mathrm{pairs}} - P) = (N_{\mathrm{pairs}} - (\ell-1)) - (N_{\mathrm{pairs}} - P) = 1 + P - V \;,
\label{eq:number_of_loops}
\end{equation}
which is the familiar QFT relation between loops, propagators and vertices.

We finally need to acknowledge that our choice of $\vqi_\ell$ as the center of our new coordinate system was arbitrary. Any other choice would still lead to a set of generalised Fourier
momenta like in \eref{eq:generalised_fourier_momenta} which obey the conservation law \eref{eq:fourier_momentum_conservation} and allow to integrate out momenta associated with open pairs.
We therefore can always reconstruct the appropriate set of Fourier momenta directly from the conservation law by following the Feynman rules of section \ref{sec:feynman_rules}.

\section{Time flow loops} \label{app:time_loops}

We need to show that any loop-subdiagram in the time-flow diagram language of \eref{eq:response_factor_diagrams} leads to a factor of zero.
In such a loop, each dot corresponding to a field label $\bar{u}$ must be the origin of an arrow line, thus there must be a response factor $b_j(\bar{u})$ and $\bar{u}$ is therefore a $\Phi_B$-label.
In a compound cumulant \eref{eq:compound_cumulant} such labels are always paired with a $\sigma$-matrix element which forces $\vl_{\bar{u}} = 0$. The phase-shift vectors
corresponding to the arrow lines in the loop thus reduce to
\begin{equation}
 \mytikz{\hortwopattern[t_{\bar{u}}][t_{\bar{r}}] \arrline{p1}{p2} } = \ii \, \vec{L}_{p,\bar{r}}(t_{\bar{u}}) \cdot \vk_{\bar{u}} \bigg|_{\vl_{\bar{r}}=0} = \gqp(t_{\bar{r}},t_{\bar{u}}) \, \vk_{\bar{r}} \cdot \vk_{\bar{u}} \propto \heavi\left( t_{\bar{r}} - t_{\bar{u}} \right)\;.
\label{eq:reduced_time_flow_arrow}
\end{equation}
A general closed loop of $k$ dots has the form
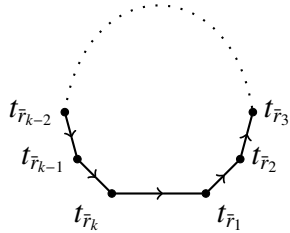
\begin{center}
\tikz[baseline=4.0ex]{
  \coordinate (center) at (0,0);
  \newlength{\radius}
  \setlength{\radius}{1.25cm}  
  \path (center) ++ (-60:\radius) coordinate[pdot] (t1);
  \node[below right] at (t1) {$t_{\bar{r}_1}$};  
  \path (center) ++ (-30:\radius) coordinate[pdot] (t2);
  \node[right] at (t2) {$t_{\bar{r}_2}$};  
  \path (center) ++ (0:\radius) coordinate[pdot] (t3);
  \node[right] at (t3) {$t_{\bar{r}_3}$};  
  \path (center) ++ (180:\radius) coordinate[pdot] (tm2);
  \node[left] at (tm2) {$t_{\bar{r}_{k-2}}$};  
  \path (center) ++ (210:\radius) coordinate[pdot] (tm1);
  \node[left] at (tm1) {$t_{\bar{r}_{k-1}}$};  
  \path (center) ++ (240:\radius) coordinate[pdot] (tm);
  \node[below left] at (tm) {$t_{\bar{r}_k}$};  
  \arrline{t1}{t2} \arrline{t2}{t3} \arrline{tm2}{tm1} \arrline{tm1}{tm} \arrline{tm}{t1};
  \draw[loosely dotted, thick] (t3) .. controls (0.75\radius,1.5\radius) and (-0.75\radius,1.5\radius) .. (tm2);
  }
\end{center}
\begin{eqnarray}
 &\heavi(t_{\bar{r}_{k}}-t_{\bar{r}_{k-1}}) \, \heavi(t_{\bar{r}_{k-1}} - t_{\bar{r}_{k-2}}) \ldots \heavi(t_{\bar{r}_2}-t_{\bar{r}_1}) \, \heavi(t_{\bar{r}_1}-t_{\bar{r}_k}) \nonumber \\
 \Rightarrow& t_{\bar{r}_k} \geq  t_{\bar{r}_{k-1}} \geq \ldots \geq t_{\bar{r}_1} \geq t_{\bar{r}_k} \;. \nonumber 
\end{eqnarray}
This time ordering is only possible if all instances of time are the same $t_{\bar{r}_1} = \ldots = t_{\bar{r}_k}$. 
But then all arrow lines in the loop are proportional to $\gqp(t_{\bar{r}_1},t_{\bar{r}_1}) = 0$ according to \eref{eq:reduced_time_flow_arrow}, therefore the above loop contains $k$
factors of zero and must thus vanish.

We also note that if we integrate out the momentum space information from the collective fields, \ie we only work in terms of the spatial particle density $\Phi_\rho$ as in \cite{Bartelmann2016},
the shift vectors $\vec{L}_{p,r}$ have the form \eref{eq:reduced_time_flow_arrow} even before contracting mixed cumulants with the $\sigma$-matrix.
In this case the causality restriction applies directly to the bare mixed cumulants.


\bibliographystyle{iopart-num}
\bibliography{Bibliography}
\end{document}